\documentclass[a4paper, 10pt]{article}
\usepackage[utf8]{inputenc}
\usepackage[english]{babel}
\usepackage[left=2.5cm, right=2.5cm, top=2cm]{geometry}

\usepackage{hyperref}
\hypersetup{colorlinks=true,linkcolor=red,citecolor=blue,filecolor=black,urlcolor=black,
pdfauthor={Jacopo Mazza}}

\usepackage{graphicx}
\usepackage[dvipsnames]{xcolor}
\usepackage{caption}
\usepackage{subcaption}

\usepackage{siunitx}
\let\svqty\qty
\usepackage{physics}
\let\qty\svqty
\usepackage{amsmath,amssymb,amsfonts}
\def\be#1\ee{\begin{align}#1\end{align}} 
\def\bse#1\ese{\begin{subequations}#1\end{subequations}}

\newcommand{\Lie}[2]{\ensuremath{\pounds_{#1} #2 }} 
\newcommand{\iu}{\ensuremath{\mathrm{i}}} 

\usepackage[capitalize]{cleveref}
\Crefname{equation}{Eq.}{Eqs.}
\Crefname{figure}{Fig.}{Fig.s}
\Crefname{table}{table}{tables}
\crefname{subequation}{Eqs.}{Eqs.}
\labelcrefformat{subequation}{#2(#1)#3}
\crefformat{subequations}{#2(#1)#3}
\crefname{section}{Sec.}{Secs.}

\usepackage{comment}
\usepackage{soul}

\newcommand{\qu}[1]{``#1''} 

\usepackage{authblk}
\newcommand{\IJCLab}{\affil{Université Paris--Saclay, CNRS/IN2P3, IJCLab, 91405 Orsay, France}}

\begin{document}

\title{Properties of general stationary axisymmetric spacetimes: circularity and beyond}
\author{Eugeny Babichev}
\author{Jacopo Mazza}
\IJCLab
\date{}

\maketitle
\begin{abstract}
    We analyse properties of general stationary and axisymmetric spacetimes, with a particular focus on circularity --- an accidental symmetry enjoyed by the Kerr metric, and therefore widely assumed when searching for rotating black hole solutions in alternative theories of gravity as well as when constructing models of Kerr mimickers.
    Within a gauge specified by seven (or six) free functions, the local existence of which we prove, we solve the differential circularity conditions and translate them into algebraic relations among the metric components.
    This result opens the way to investigating the consequences of circularity breaking in a controlled manner.
    In particular, we construct two simple analytical examples of non-circular deformations of the Kerr spacetime. 
    The first one is \qu{minimal}, since the horizon and the static limit are identical to their Kerr counterparts, except for the fact that the horizon is not Killing and its surface gravity is therefore not constant.
    The second is \qu{not so minimal}, as the horizon's profile can be chosen arbitrarily and the difference between the horizon and the so-called rotosurface can be appreciated.
    Our findings thus pave the way for further research into the phenomenology of non-circular stationary and axisymmetric spacetimes.
\end{abstract}

\section[Introduction]{\label{sec:Intro}Introduction}

Modern instruments and methods allow to observe black holes and other compact objects with unprecedented accuracy. 
Although horizons cannot be observed directly, there exist indirect ways to infer the presence of a black hole.
Examples include:
the gravitational waves from mergers of binaries~\cite{LIGOScientific:2016aoc,LIGOScientific:2017vwq};
the images of the electromagnetic emission from accretion disks~\cite{EventHorizonTelescope:2020qrl};
and the trajectories of stars around the supermassive galactic centre~\cite{GRAVITY:2018ofz}.
Such observations serve as an excellent laboratory for assessing the validity of General Relativity (GR), which so far passed all tests. 
In the future, the precision of observations will only increase, as new and updated instruments --- such as LISA and the Einstein Telescope --- will become operational.
However, it is widely believed that GR cannot be the ultimate theory of gravity, for various reasons --- from its lack of perturbative renormalisability to the issues in cosmology, such as the unsolved puzzle of dark energy.
The question thus arises of how to search for deviations from GR, having present and future observations of black holes in mind. 
From the theoretical point of view, one expects to find examples of metrics that deviate from the Kerr solution, and thus allow for non-null tests of GR and of the Kerr paradigm.

There are two different approaches for introducing deviations from GR black hole solutions. 
The direct approach consists in taking a particular modified gravity theory and searching for black hole solutions of the theory. 
Although there are already quite a number of numerical and even analytic spherically symmetric solutions in modified gravity models (for review see e.g.~\cite{Babichev:2016rlq,Lehebel:2018zga,Babichev:2023psy}), only a few exact rotating solutions are known due to the high complexity of the equations of motion.
Apart from stealth solutions, i.e.~those in which the metric is that of Kerr(--de Sitter)~\cite{Charmousis:2019vnf,FranzinKerrBlack2024}, to our best knowledge there is only one known exact non-stealth black hole solution: 
the disformal Kerr solution in generalised scalar--tensor gravity~\cite{AnsonDisformingKerr2021,BenAchour:2020fgy} (for a counterpart of such solutions in scalar--vector theories, see~\cite{MinamitsujiDisformalTransformation2020};one can also mention the metric constructed in conformal gravity~\cite{Astorino:2013sfa}, although the presented solution is pathological due to the vanishing of the Einstein-Hilbert term.). 
The advantage of this approach to look for deviations from GR is that a solution allows to study all the properties of black holes, including those that are related to gravitational degrees of freedom, since the theory is given. 
Indeed, various observational consequences of the disformed Kerr solution have been studied in a series of papers~\cite{AnsonDeformedBlack2021,BabichevTestingDisformal2024a,TakamoriTestingNoncircularity2023,ChenTestingGravity2021}.
A clear disadvantage is, however, that for the moment there are few available non-trivial exact solution to investigate. 

To overcome this difficulty, an alternative approach is to be agnostic and to construct an \emph{ad hoc} metric that describes a deviation from the Kerr metric. 
Various parametrisations of Kerr deviations have been proposed in the past, see e.g.~\cite{JohannsenMetricRapidly2011,CardosoGenericParametrizations2014,RezzollaNewParametrization2014,KonoplyaGeneralParametrization2016,HeumannIdentifyingEvent2023}. 
The advantage of this method is that one can construct more general alternatives to the Kerr metric, which allow to look for smoking guns of GR modifications. 
On the other hand, an \emph{ad hoc} metric, strictly speaking, makes it impossible to solve dynamical problems, such as computing quasinormal modes or extreme mass ratio inspirals, without extra assumptions. 
Besides, many examples of \emph{ad hoc} Kerr deviations have been shown to be pathological, because of naked singularities or non-causal spacetimes~\cite{Johannsen:2013rqa}. 
This problem can be often attributed to the method by which these examples are constructed, in particular when a deformation is built using coordinates that are singular at the horizon of the seed Kerr metric, which normally leads to a naked singularity for the deformation. 
This issue can be handled, however, by means of horizon-penetrating coordinates, which are explicitly not singular at the horizon.

The majority of Kerr deformations suggested in the literature remain circular, as in the case of the Kerr metric. 
The circularity property of a metric is deeply related to the separability of the equations of motion of geodesics~\cite{BenentiRemarksCertain1979}: separability of geodesics implies circularity.
It is therefore much simpler to analyse the motion of particles and light rays in a circular metric. 
On the other hand, although vacuum GR solutions only support circular spacetimes, even in GR non-vacuum solutions may have a non-circular metric~\cite{GourgoulhonNoncircularAxisymmetric1993}.
In modified gravity, the study of black hole solutions strongly suggest that the lack of circularity is a generic feature. 
One can name in this context the already mentioned disformed Kerr solution~\cite{AnsonDisformingKerr2021}, as well as the numerically constructed rotating black holes in Einstein-\ae{}ther theory~\cite{adam_rotating_2021} and in semiclassical gravity~\cite{FernandesRotatingBlack2023}.
Therefore a natural question would be to ask what is the general non-circular extension of Kerr deformations. 
Recently there has been some progress in this direction~\cite{GhoshProbingSpacetime2024,GhoshParameterizedNoncircular2024,DelaporteParameterizationsBlackhole2022a,eichhorn_image_2021,eichhorn_locality-principle_2021}.
Ref.~\cite{GhoshParameterizedNoncircular2024} assumes the Kerr metric in Boyer-Lindquist coordinates and constructs a non-circular metric with a small parametrised deviation from Kerr.
In order to avoid the above-mentioned problem that generically arises when deforming the metric in singular (on the horizon) coordinates, Ref.~\cite{GhoshParameterizedNoncircular2024} had to choose a specific form of deviations, besides the assumption of perturbative deviations. 
In this respect, the deviations suggested in~\cite{GhoshParameterizedNoncircular2024} are not general. 
On the other hand, the authors of~\cite{DelaporteParameterizationsBlackhole2022a} used the horizon-penetrating coordinates of the Kerr metric and suggested a general Ansatz for non-circular deviations of the metric.
However, Ref.~\cite{DelaporteParameterizationsBlackhole2022a} did not manage to show that this particular Ansatz is the most general one. 
In addition, due to the use of the horizon-penetrating coordinates, the conditions for circularity are not obvious to formulate in this approach. 

In the present paper we take a step forward in the study of the most general non-circular stationary metric with axial symmetry. 
We show how, using the gauge freedom, the most general such metric can be brought (at least locally) to a Kerr-like form, with particular components of the metric set to zero.
The choice of coordinates corresponding to this metric are analogous to the Kerr penetrating coordinates, and in fact coincide with those in the case of GR.
We then analyse the differential conditions that enforce circularity, and solve them in terms of algebraic relations among the components of the  metric. 
This allows us to construct particularly simple examples of non-circular metrics that can be used to identify key properties of the presence of non-circularity.

\section{\label{sec:Circ}Circularity vs.~non-circularity}

A rather rich introduction to the topic of circularity can be found in the excellent lecture notes by Carter \cite{CarterBlackHoles1973,CarterRepublicationBlack2010}.\footnote{Along with scores of insightful physics, these notes contain an interesting historical quirk. They state that a consensus had been reached at the 1972 Les Houches School on the best French translation of \qu{black hole}: \textit{\qu{piège noir}}. Evidently, this proposal has not caught on, as the only term used in current French appears to be \textit{\qu{trou noir}}.}
Following their reasoning, we may identify (at least) two logically distinct but largely equivalent notions of circularity.
Although they differ slightly in their rationale, both rely on the existence of two Killing vectors and therefore apply to stationary and axisymmetric spacetimes.
Here and throughout, we call $\xi^\mu$ and $\psi^\mu$ the Killing vectors associated to stationarity and axisymmetry, respectively --- with the understanding that $\xi^\mu$ is timelike in a neighbourhood of infinity, while $\psi^\mu$ is everywhere spacelike, has closed orbits, and vanishes on the axis of symmetry.

The first and probably most popular of such notions refers to a geometric property of the Killing flows, and therefore of the spacetime metric.
Specifically, a spacetime is said to be circular if the two-dimensional surfaces generated by the action of the Killing isometries are everywhere orthogonal to a family of codimension-two surfaces.
Frobenius' theorem allows to translate the requirement that a spacetime be circular into the following two equations involving the Killing fields $\xi^\mu$ and $\psi^\mu$:
\bse\label{eq:CircDef}
\be
\left( \xi_\mu \dd{x^\mu}\right) \wedge \left(\psi_\nu \dd{x^\nu} \right) \wedge \dd{\left(\xi_\rho \dd{x^\rho} \right) } &= 0 \, ,\\
\left( \psi_\mu \dd{x^\mu}\right) \wedge \left(\xi_\nu \dd{x^\nu} \right) \wedge \dd{\left(\psi_\rho \dd{x^\rho} \right) } &= 0\, .
\ee
\ese
These equations, to which we will often refer as \emph{circularity conditions}, are meant to hold everywhere in the spacetime.

The second notion, instead, deals with a generic flow, specified by a vector field $u^\mu$.
Such vector might represent, for instance, the velocity of a particle or of a cloud of gas, or again an electromagnetic current.
Hence, compared to the alternative above, this second notion refers more directly to the properties of matter.
Specifically, the vector $u^\mu$ is said to be circular if it can be written as a linear combination of the Killing vectors $\xi^\mu$ and $\psi^\mu$.
Note that the coefficients of such combination need not be constant, so a circular vector is not necessarily Killing. 
The requirement that $u^\mu$ be circular may be written as 
\be\label{eq:CircDefFlow}
u^{[ \mu} \xi^\nu \psi^{\rho ]} = 0 \,.
\ee

The connection between these two notions emerges when one specifies the previous definition to a particular flow vector: namely, one that is an eigenvector of the Ricci tensor. 
Indeed, if $u^\mu$ is such that
\be
R^\mu _{\ \nu}u^\nu = \lambda u^\mu
\ee
for some scalar $\lambda$, circularity in the sense of \cref{eq:CircDefFlow} is equivalent to 
\be
u^\mu R_{\mu [\nu} \xi_\rho \psi_{\sigma ]} = 0\, ,
\ee
which in turn implies the following two equations:
\bse\label{eq:CircDefRicci}
\be
\xi^\mu R_{\mu [\nu} \xi_\rho \psi_{\sigma ]} &= 0 \, ,\\
\psi^\mu R_{\mu [\nu} \xi_\rho \psi_{\sigma ]} &= 0 \, .
\ee\ese
(Carter \cite{CarterBlackHoles1973,CarterRepublicationBlack2010} states that the implication holds in the opposite direction too, i.e.~that \cref{eq:CircDefRicci} is necessary and sufficient for the eigenvector to be circular; however, we have not been able to reproduce this result.)
These last equations entail that the vectors $R_{\mu \nu}\xi^\nu$ and $R_{\mu \nu}\psi^\nu$ are themselves circular as per \cref{eq:CircDefFlow}, a property sometimes known as \emph{invertibility} of the Ricci tensor \cite{KundtOrthogonalDecomposition1966,CarterKillingHorizons1969,CarterBlackHoles1973,CarterRepublicationBlack2010,Xie:2021bur}.
By exploiting the Killing equation(s) and some algebraic identities, one can prove that \cref{eq:CircDef} imply \cref{eq:CircDefRicci} --- see \cite{CarterBlackHoles1973,CarterRepublicationBlack2010} for details.
Moreover, assuming that \cref{eq:CircDef} are satisfied at least at one point, one can also prove that \cref{eq:CircDefRicci} imply \cref{eq:CircDef} --- a result known as Papapetrou's theorem \cite{PapapetrouChampsGravitationnels1966}.
Since, by definition, $\psi^\mu$ vanishes on the axis of symmetry, \cref{eq:CircDef} are always satisfied there, hence \cref{eq:CircDef} and \cref{eq:CircDefRicci} are practically equivalent in standard situations.

The motivation for focusing on an eigenvector of the Ricci tensor stems primarily from Einstein's equations, since, when these hold, eigenvectors of the Ricci are also eigenvectors of the energy--momentum tensor.
Moreover, Einstein's equation can be used to translate \cref{eq:CircDefRicci} into the statement that non-circular energy--momentum fluxes be absent.
Hence, the equivalence of the two notions of circularity exemplifies the tight relationship that exists in GR between matter and geometry --- or, more precisely, between the symmetries of the sources and those of the metric.
One may infer that in GR, roughly speaking, circular sources give rise to circular spacetimes.
For instance, in the case of a perfect fluid, the circularity of the fluid's velocity is sufficient to deduce the circularity of the metric;
the same is true in the case of an electromagnetic field, when circularity of the electromagnetic current is assumed.
Note that GR vacuum solutions, with or without a cosmological constant, as well as all Ricci-flat configurations, trivially satisfy \cref{eq:CircDefRicci} and therefore are automatically circular.

Clearly, many cases of physical interest will require sacrificing some degree of symmetry but, all in all, circularity is a rather common feature of GR solutions --- particularly of those that describe black holes. 
It is immediate to suspect that things might become more nuanced as one moves away from GR.
Nonetheless, circularity remains a very common assumption when constructing alternatives to the Kerr metric, or when searching for stationary and axisymmetric solutions in alternative theories of gravity.
In the following, we will first comment on why this is the case, by providing several good arguments in favour of circularity;
then, we will discuss why restricting to such assumption might be too strong a limitation.

\subsection{\label{subsec:WhyCirc} Why circularity? }

A rather straightforward motivation for considering circular metrics even beyond GR is that some solutions to alternative theories of gravity are, in fact, circular.
This is clearly the case for so-called stealth solutions, whereby the metric is that of Kerr (e.g.~\cite{Charmousis:2019vnf,FranzinKerrBlack2024}), but other, less trivial examples exist --- such as the numerical solutions in dilatonic Einstein--Gauss--Bonnet theory~\cite{Kleihaus:2011tg,Kleihaus:2015aje}, as well as perturbative rotating solutions in scalar-Einstein-Gauss-Bonnet gravity and dynamical Chern-Simons gravity~\cite{Nakashi:2020phm}.
Moreover, Ref.~\cite{Xie:2021bur} put forward a strong argument according to which solutions in generic effective field theories of gravity that are perturbatively connected to GR must be circular.
Clearly, this point is very relevant to our discussion and we thus return to it in the next subsection, with the aim of clarifying why this conclusion is often circumvented in practice.

When it comes to constructing models of Kerr mimickers, however, probably the strongest argument in favour of circularity is the dramatic simplification that this brings about.
As mentioned, the circularity conditions of \cref{eq:CircDef} ensure the existence of a foliation in terms of two-dimensional surfaces everywhere orthogonal to the Killing vectors.
Such surfaces are usually dubbed \emph{meridional surfaces}, while those that are generated by the Killing flows are \emph{surfaces of transitivity}.
In a circular spacetime, one can therefore adapt the choice of coordinates to such foliation, by picking two coordinates to chart the meridional surfaces and two other coordinates to chart the surfaces of transitivity (cf.~\cite{BokulicGeneralizationsChallenges2023}). 

In such coordinates, the metric is automatically block diagonal, since it consists in a $2\times2$ \qu{Killing} block and a $2\times2$ \qu{meridional} block, while all other components are zero.
Typically, one can also diagonalise the \qu{meridional} block, so that the metric has only five non-vanishing components: four components on the diagonal and only one off of it.
In the Kerr spacetime, a coordinate system of this kind is the familiar Boyer--Lindquist one, with $t_\text{BL}$ and $\phi_\text{BL}$ being associated to the Killing vectors, and $r_\text{BL}$ and $\theta_\text{BL}$ charting the meridional surfaces.
For this reason, we will say that a circular metric can be brought to Boyer--Lindquist form --- or, equivalently, that a circular spacetime admits Boyer--Lindquist-like coordinates.

Such technical simplification opens the way to establishing a series of physically relevant properties.
For example, a circular metric expressed in adapted coordinates is automatically symmetric under the simultaneous sign flips $\xi^\mu \mapsto - \xi^\mu$ and $\psi^\mu \mapsto - \psi^\mu$ --- a property sometimes known as \qu{$t$--$\phi$ symmetry}. 
Moreover, one can prove that in a circular spacetime the so-called rotosurface (cf.~\cref{subsubsec:Surfaces}), i.e.~the locus of points at which the linear combination $\xi^\mu + \Omega\, \psi^\mu$ becomes null, necessarily coincides with the black hole's event horizon.\footnote{To get a deeper grasp of this result, one may draw an analogy with the more familiar case of staticity, defined by the requirement that the Killing vector $\xi^\mu$ be hypersurface orthogonal. In that case, one similarly finds that the ergosurface coincides with the (Killing) horizon. This analogy is actually quite profound --- once again, we refer to \cite{CarterBlackHoles1973,CarterRepublicationBlack2010} for further details.}
Here, $\Omega := - (\xi_\mu\, \psi^\mu)/(\psi_\nu \, \psi^\nu)$ is the angular velocity of frame dragging, and one can additionally prove that if the spacetime is circular then $\Omega$ is a constant over the event horizon, which is therefore also a Killing horizon.
This fact is extremely consequential e.g.~for the notion of surface gravity \cite{CroppSurfaceGravities2013,BainesExplicitFormulae2023a}.
For further details, we refer again to \cite{CarterBlackHoles1973,CarterRepublicationBlack2010}.

We close this short summary of enticing features of circular spacetimes by mentioning that circularity is a necessary (albeit not sufficient) condition for the additive separation of variables in the Hamilton--Jacobi equation that describes geodesics --- a point we will return to in \cref{subsubsec:Geodesics}.
Such separability is associated to the existence of a rank-two Killing tensor, which gives rise to an additional conserved quantity.
The most general metric admitting such Killing tensor is known \cite{BenentiRemarksCertain1979,JohannsenRegularBlack2013} and, as one can verify directly, it happens to be circular \cite{DelaporteParameterizationsBlackhole2022a}.
Note that the existence of a Killing tensor is necessary (but technically not sufficient) also for the multiplicative separability of the Klein--Gordon equation \cite{FrolovBlackHoles2017,GiorgiCarterTensor2024,BainesKillingTensor2021}.
Hence, circular spacetimes are especially suited for phenomenological investigations, as the dynamics of test particles and fields can be solved with relative ease.

\subsection{\label{subsec:WhyNotCirc} Why \emph{not} circularity? }

As it was mentioned above, Ref.~\cite{Xie:2021bur} claims that stationary and axisymmetric black hole solutions in effective field theories of gravity are circular, provided the theory and the solution in question are continuously connected to GR and to one of its solutions.
At the same time, in the Introduction, \cref{sec:Intro}, we have listed several examples of non-circular solutions in modified gravity, and argued that the lack of circularity is a generic feature in this context.
To shed light on this apparent conundrum, we thus retrace the proof of~\cite{Xie:2021bur} and highlight when and why it might not apply.

Schematically, the argument of \cite{Xie:2021bur} concerns theories with Lagrangian
\be\label{eq:LEFT}
\mathcal{L} = \mathcal{L}_0 + \alpha \mathcal{L}_M 
\ee
depending on the spacetime metric and on additional fields collectively denoted $\varphi$.
Here, $\mathcal{L}_0$ contains the Ricci scalar and all operators of order up to four, while $\alpha$ is a coupling constant that suppresses the higher-order operators contained in $\mathcal{L}_M$.
The equations of motion deriving from \eqref{eq:LEFT} are assumed to admit a power series solution in $\alpha$.
The authors of \cite{Xie:2021bur} proved that if $\varphi$ satisfies a suitable symmetry requirement, then (\textit{i}) the spacetime is circular at zeroth order in $\alpha$, and (\textit{ii}) this property can be extended order by order in $\alpha$.
The precise symmetry requirement depends on the rank of $\varphi$ as a tensor:
if it is a scalar, one needs $\xi^\mu \partial_\mu \varphi = \psi^\mu \partial_\mu \varphi = 0$;
if it is a vector, one needs it to be circular in the sense of \cref{eq:CircDefFlow};
and so on for higher ranks.
We stress that this symmetry requirement is absolutely crucial for the entire argument.

It is important to stress that the results of~\cite{Xie:2021bur} have their limitations.
A first observation is that the perturbative argument might fail, for instance if the solution does not reduce to a GR solution in the limit $\alpha \to 0$.
A second objection, more relevant to the present discussion, is that the symmetry requirement on $\varphi$ might not always be justified.
To see why, note that in the case of a scalar field $\xi^\mu \partial_\mu \varphi = \psi^\mu \partial_\mu \varphi = 0$ simply expresses the fact that the scalar itself should be stationary and axisymmetric, and this condition is therefore perfectly logical;
however, the equivalent conditions for higher-rank fields do not descend directly from the spacetime symmetries and are therefore genuinely independent.
Hence, the decision on whether to impose them or not is arbitrary.

A good example of this fact is provided by shift-symmetric scalar--tensor theories, in which the Lagrangian depends on the scalar only through its gradient.
In this context, one might still want to impose $\psi^\mu \partial_\mu \varphi = 0$, as this prevents the scalar from being multivalued.
However, $\xi^\mu \partial_\mu \varphi = 0$ is not strictly required, as one can have stationary solutions even with a less stringent condition.
In fact, one typically imposes $\Lie{\xi}{\left( \partial_\mu \varphi \right)} =0$, which allows a linear time dependence of $\varphi$ on the Killing time, i.e.~$\varphi=\text{const}\times t+\psi(r)$ as suggested in~\cite{Babichev:2013cya}.
To fall within the assumptions of \cite{Xie:2021bur}, one would need to impose either $\xi^\mu \partial_\mu \varphi = 0$ or $\partial_\mu \varphi \propto \xi_\mu - \Omega\, \psi_\mu$.
Both these conditions restrict the space of solutions, more than the combination of stationarity and axisymmetry does.
Hence, one can expect non-circular solutions to exist along with circular ones.
Indeed, the disformed Kerr metric of \cite{AnsonDisformingKerr2021} provides an example of a non-circular solution in a shift-symmetric theory. 

Another excellent example is provided by Einstein--\ae{}ther theory, a vector--tensor theory characterised by a constraint forcing the \ae{}ther vector $u^\mu$ to be everywhere timelike and of unit norm.
In this case too one usually assumes $u_\mu\, \psi^\mu = 0$, for reasons pertaining to the peculiar notion of causality inherent to this theory.
However, as already hinted to in \cite{Xie:2021bur}, asking that $u_\mu$ be circular is at odds with the unit-norm constraint and overall rather \qu{unnatural} in this context.
For instance, this assumption automatically excludes the possibility of having universal horizons \cite{PorroUniversalHorizons2025}. 
Hence, one generically expects this theory to evade the conclusions of \cite{Xie:2021bur}.
Of course, this does not imply that stationary and axisymmetric Einstein--\ae{}ther black holes are \emph{necessarily} non-circular.
In fact, a (circular) stealth Kerr solution is known to exist in a corner of the parameter space \cite{FranzinKerrBlack2024}.
However, one cannot expect circularity to hold in general. 
Thus, it is not surprising that the rotating black holes recently found numerically in \cite{adam_rotating_2021} fail to be circular.

These examples demonstrate that assuming circularity \emph{a priori} is often not justified and can lead to unnecessarily stringent conclusions.
This is true when writing Ans\"{a}tze for stationary and axisymmetric solutions in alternative theories, and even more so when constructing phenomenological models of black hole mimickers.
This motivates us to consider general stationary and axisymmetric metrics, without assuming circularity, and to investigate the consequences of circularity breaking.

\section{\label{sec:Gauge}General axisymmetric stationary metric and gauge choice}

\subsection{\label{subsec:Ansatz}Gauge choices for the metric}

Since we wish to study stationary and axially symmetric spacetimes, it appears most convenient to choose coordinates adapted to these symmetries.
We thus introduce the coordinates $v$ and $\phi$ along the Killing orbits, such that
\be\label{eq:KillCoord}
\xi^\mu \partial_\mu = \partial_v 
\qq{and}
\psi^\mu \partial_\mu = \partial_\phi\, ;
\ee
and complement them with two other coordinates $r$ and $\theta$, so that
\be
\label{eq:gCoord}
g_{\mu \nu} \left(x^\alpha\right) = g_{\mu \nu}(r,\theta) \, .
\ee
Such coordinates are not uniquely determined, since any diffeomorphism of the form
\be\label{eq:diffeo}
\begin{split}
\tilde{v} &= v + V \left(r,\theta \right) \\
\tilde{r} &= R\left(r,\theta \right) \\
\tilde{\theta} &= \Theta \left(r,\theta \right)\\
\tilde{\phi} &= \phi + \Phi \left(r,\theta \right)
\end{split}
\ee
yields new coordinates that satisfy the same properties as in \cref{eq:KillCoord,eq:gCoord}.

These coordinates are inspired by the Kerr (ingoing or outgoing) coordinates, hence we implicitly assume they are defined on the usual domains and have the familiar interpretation.
Namely, $\phi \in [0,2\pi)$ and $\theta \in [0,\pi]$ are angles, with $\theta = 0$ and $\theta = \pi$ both identifying the axis of symmetry.
Moreover, $v \to \pm \infty$ corresponds to the future and past infinity, respectively, while $r \to \infty$ identifies the asymptotically faraway region.
In addition, we wish to ensure that the coordinates are horizon penetrating whenever a horizon is present.
The simplest way to do this is to require that the metric components be manifestly regular outside and at the event horizon. 

So far, our assumptions are therefore very minimal and the metric is correspondingly general.
In particular, all of its components are generically not zero.
Luckily, we can still exploit the freedom encoded in \cref{eq:diffeo} to simplify the form of the metric.
A particularly convenient choice, which we make in analogy with the Kerr coordinates, consists in setting to zero all the mixed $\theta$ components, i.e.~taking
\be\label{ansatz0}
g_{\mu \theta} \propto \delta^\theta_{\ \mu}\, .
\ee
We shall refer to this choice as the \emph{orthogonal gauge}, since it entails that the coordinate basis vector $\partial_\theta$ is orthogonal to $\partial_v$, $\partial_r$, and $\partial_\phi$.
Note that, according to \cref{ansatz0}, the mixed $\theta$ components of the inverse metric are similarly zero:
\be
g^{\mu \theta} \propto \delta^\mu_{\ \theta}\, .
\ee
Additionally, again in analogy with the Kerr coordinates, we may demand that $v$ be a null coordinate, in the sense that the hypersurfaces $v = \text{const.}$ be null.
This requirement is implemented by setting
\be\label{eq:KerrLike}
g_{rr} = 0\, .
\ee
We shall refer to the gauge specified by the combination of \cref{ansatz0,eq:grrAnsatz} as \emph{Kerr-like gauge}.
In passing, we point out that the Kerr-like gauge is equivalent to the Ansatz proposed in \cite{DelaporteParameterizationsBlackhole2022a}.

To summarise, the metric we consider is
\be\label{eq:OrthoGauge}
\dd{s}^2 &= g_{vv} \dd{v}^2 + g_{\theta\theta} \dd{\theta}^2 + g_{rr} \dd{r}^2 + g_{\phi \phi} \dd{\phi}^2 \nonumber\\
&\phantom{=} + 2 g_{v\phi} \dd{v} \dd{\phi} + 2 g_{vr} \dd{v} \dd{r} + 2 g_{r\phi} \dd{r} \dd{\phi}\, ,
\ee
where all the metric components depend only on $r$ and $\theta$, which we refer to as orthogonal gauge.
We further call Kerr-like gauge \cref{eq:OrthoGauge} with $g_{rr} = 0$.

\bigskip

In the next subsections, we shall first of all prove that such a gauge choice is allowed. 
Specifically, we will show the local existence of a coordinate transformation that brings an arbitrary stationary and axisymmetric metric into the Kerr-like gauge;
the limits of applicability of our reasoning will thus become clear as we spell it out. 
Then, we shall characterise these metrics by describing some of their features: 
we will present three relevant surfaces (static limit, rotosurface, and event horizon); construct a null \qu{Kinnersley-like} tetrad; and finally make a few comments on geodesics.
This choice of features is clearly partial, yet functional to the rest of the discussion, and definitely leaves room for further investigations.
At each step in the presentation, we shall make an effort to state which gauge --- orthogonal or full Kerr-like --- is strictly necessary for the particular conclusion under consideration.

\subsection{\label{subsec:Proof}Proof of local existence of the gauge choice}

To prove that the Kerr-like gauge introduced above does in fact exist, we start from a generic metric $g_{\tilde{\mu}\tilde{\nu}} (\tilde{r}, \tilde{\theta})$ and perform the change of coordinates~\cref{eq:diffeo} to a $g_{\mu \nu}(r,\theta)$ which we take in the Kerr-like form of \cref{eq:OrthoGauge} with $g_{rr}=0$.
We get the following three equations, corresponding to the orthogonal gauge alone
\be\label{eq:gutheta}
0 = g_{v\theta} &= g_{\tilde{v}\tilde{v}}\pdv{V}{\theta}+g_{\tilde{v}\tilde{r}}\pdv{R}{\theta}+g_{\tilde{v}\tilde{\theta}}\pdv{\Theta}{\theta}+g_{\tilde{v}\tilde{\phi}}\pdv{\Phi}{\theta}\, , \\ 
\label{eq:gphitheta}
0 = g_{\phi \theta} &= g_{\tilde{\phi}\tilde{v}}\pdv{V}{\theta}+g_{\tilde{\phi}\tilde{r}}\pdv{R}{\theta}+g_{\tilde{\phi}\tilde{\theta}}\pdv{\Theta}{\theta}+g_{\tilde{\phi}\tilde{\phi}}\pdv{\Phi}{\theta}\, , \\
\label{eq:grtheta}
0 = g_{r \theta} &= \left[ \text{\cref{eq:gutheta}} \right] \pdv{V}{r} + \left[ \text{\cref{eq:gphitheta}} \right] \pdv{\Phi}{r} \nonumber\\
&\phantom{=} + \left[ g_{\tilde{v} \tilde{r} } \pdv{R}{r} + g_{\tilde{v} \tilde{\theta} } \pdv{\Theta}{r} \right]\pdv{V}{\theta} 
+ \left[ g_{\tilde{r} \tilde{\phi} } \pdv{R}{r} + g_{\tilde{\theta} \tilde{\phi} } \pdv{\Theta}{r} \right]\pdv{\Phi}{\theta} \nonumber\\
&\phantom{=}
+  \left[ g_{\tilde{r} \tilde{r} } \pdv{R}{r} + g_{\tilde{r} \tilde{\theta} } \pdv{\Theta}{r} \right]\pdv{R}{\theta} 
+ \left[ g_{\tilde{r} \tilde{\theta}} \pdv{R}{r} + g_{\tilde{\theta} \tilde{\theta}} \pdv{\Theta}{r} \right]\pdv{\Theta}{\theta}\, ;
\ee
and the following fourth equation further enforcing the full Kerr-like gauge
\be\label{eq:grr}
0 = g_{rr} &= g_{\tilde{v} \tilde{v}}\left(\pdv{V}{r}\right)^2 + g_{\tilde{r} \tilde{r}}\left(\pdv{R}{r}\right)^2 + g_{\tilde{\theta} \tilde{\theta}}\left(\pdv{\Theta}{r}\right)^2 + g_{\tilde{\phi} \tilde{\phi}}\left(\pdv{\Phi}{r}\right)^2 \nonumber\\
&\phantom{=} + 2 g_{\tilde{v} \tilde{r}} \pdv{V}{r} \pdv{R}{r} + 2 g_{\tilde{v} \tilde{\theta}} \pdv{V}{r} \pdv{\Theta}{r} + 2 g_{\tilde{v} \tilde{\phi}} \pdv{V}{r} \pdv{\Phi}{r} \nonumber \\
&\phantom{=} + 2 g_{\tilde{r} \tilde{\theta}} \pdv{R}{r} \pdv{\Theta}{r} + 2 g_{\tilde{r} \tilde{\phi}} \pdv{R}{r} \pdv{\Phi}{r} + 2 g_{\tilde{\theta} \tilde{\phi}} \pdv{\Theta}{r} \pdv{\Phi}{r} \, .
\ee
We thus have four non-linear partial differential equations of the first order, \cref{eq:gutheta,eq:grtheta,eq:gphitheta,eq:grr}, for four unknown functions $V \left(r,\theta \right)$, $R \left(r,\theta \right)$, $\Theta \left(r,\theta \right)$, and $\Phi \left(r,\theta \right)$.
The fact the system of equations is not overdetermined indicates that it can be resolved. 
Let us demonstrate this explicitly. 

The first three equations, \cref{eq:gutheta,eq:grtheta,eq:gphitheta}, contain derivatives with respect to $\theta$ and therefore they can be seen as evolution equations in the $\theta$ direction.
Assuming we can (and do) assign suitable \qu{initial} conditions on a $\theta = \text{const.}$ surface, we can thus apply the Cauchy-Kovalevskaya theorem that establishes the local existence of a solution. 
This proves the (local) existence of the orthogonal gauge.

On the other hand, \cref{eq:grr} does not contain derivatives with respect to $\theta$, and should therefore be interpreted as a constraint equation. 
We can, however, transform this equation into an evolution equation via the following trick. 
Let us change the independent variables as 
\be\label{eq:changertheta}
r\to r\, ,
\quad 
\theta \to \theta + r \, .
\ee
Note that this change of independent variables has a clear physical sense: we tilt the Cauchy hypersurface. 
With the above change we have 
\be
\pdv{U_i}{\theta} \to \pdv{U_i}{\theta}\, , 
\quad 
\pdv{U_i}{r} \to \pdv{U_i}{r} + \pdv{U_i}{\theta}\, ,
\ee
where we denoted collectively $U_i = \left\{ V,\, R,\, \Theta,\, \Phi \right\}$. 
One can see that \cref{eq:gutheta} and \cref{eq:gphitheta} do not changed under \eqref{eq:changertheta}, while both \cref{eq:grtheta,eq:grr} become quadratic in $\pdv*{U_i}{ \theta}$. 
Now, from the first two equations, \eqref{eq:gutheta} and \eqref{eq:gphitheta}, one can express explicitly two of the derivatives $\pdv*{U_i}{\theta}$ in terms of a linear combination of the other two.
Say we choose to express $\pdv*{V}{ \theta}$ and $\pdv*{R}{ \theta}$ in terms of $\pdv*{\Theta}{\theta}$ and $\pdv*{\Phi}{\theta}$ (this requires that the system of equations linear in $\pdv*{V}{ \theta}$ and $\pdv*{R}{ \theta}$ is not degenerate, which is in general true).
Then, substituting these expressions in the other two equations \eqref{eq:grtheta} and \eqref{eq:grr}, one finds two differential equations for two unknown functions. 
They can be written schematically as follows
\bse\label{eq:quadraticeqs}
\be
a_1 \left(\pdv{\Theta}{\theta}\right)^2 & +b_1\pdv{\Theta}{\theta}\pdv{\Phi}{\theta}+c_1 \left(\pdv{\Phi}{\theta}\right)^2 +d_1\pdv{\Theta}{\theta} +e_1\pdv{\Phi}{\theta}  = 0,\\
a_1 \left(\pdv{\Theta}{\theta}\right)^2 & +b_1\pdv{\Theta}{\theta}\pdv{\Phi}{\theta}+c_1 \left(\pdv{\Phi}{\theta}\right)^2 +d_2\pdv{\Theta}{\theta} +e_2\pdv{\Phi}{\theta}  + f_2  = 0,
\ee
\ese
where the coefficients $a_1$, $b_1$ and  $c_1$ are combinations of the metric coefficients only; 
$d_1$, $e_1$, $d_2$, $e_2$ depend on both the metric coefficients, and on $\pdv*{R}{r}$ and $\pdv*{\Theta}{r}$; 
while $f_2$ is constructed from the metric coefficients and derivatives of $U_i$ with respect to $r$.

Note that the quadratic parts in both \cref{eq:quadraticeqs} are the same, therefore one can combine the two to find, that, say, $\pdv*{\Theta}{\theta}$ can be expressed linearly in terms of $\pdv*{\Phi}{\theta}$. 
Substituting the obtained expression in one of \eqref{eq:quadraticeqs}, we find a quadratic equation in $\pdv*{\Phi}{\theta}$,
\be
a_2 \left(\pdv{\Phi}{\theta}\right)^2 +b_2  \left(\pdv{\Phi}{\theta}\right) +c_2 = 0 \, ,
\ee
where $a_2$, $b_2$ and $c_2$ depend on the metric coefficients and $\pdv*{U_i}{r}$.
This implies that we can resolve explicitly $\pdv*{\Phi}{ \theta}$ in terms of $\pdv*{U_i}{r}$ and the metric coefficients, by choosing one of the branches of the solution of the quadratic equation:
\be\label{eq:Phidiff}
    \frac{\partial \Phi}{\partial \theta} = F_\Phi\left(\frac{\partial U_i}{\partial r}, g_{\tilde\mu\tilde\nu}\right) \, .
\ee
 Note also, that the resulting expression $F_\Phi$ is analytic away from particular branching points. 

At this point we are ready to use the Cauchy-Kovalevskaya theorem on the (local) existence of the solution with given initial condition. 
Having found the solution for $\Phi=\Phi(r,\theta)$, one can readily get a similar differential equation on $\Theta$ using \eqref{eq:quadraticeqs},
\be\label{eq:Thetadiff}
    \frac{\partial \Theta}{\partial \theta} = F_\Theta\left(\frac{\partial U_i}{\partial r}, g_{\tilde\mu\tilde\nu}, \frac{\partial \Phi}{\partial \theta}\right) \, ,
\ee
where on the right-hand side there is also an explicit  dependence on the derivative of $\Phi$ with respect to $\theta$, which is however known. 
At this step we use again the Cauchy-Kovalevskaya theorem to find $\Theta= \Theta(r,\theta)$.
And, finally, from \eqref{eq:gutheta} and \eqref{eq:gphitheta} one finds differential equations for the remaining dependent function $V$ and $R$ in terms of $\Phi$ and $\Theta$:
\be
\begin{split}
\label{eq:VRdiff}
    \frac{\partial V}{\partial \theta} = F_V\left(\frac{\partial U_i}{\partial r}, g_{\tilde\mu\tilde\nu}, \frac{\partial \Phi}{\partial \theta}, \frac{\partial \Theta}{\partial \theta}\right)\, ,\\
    \frac{\partial R}{\partial \theta} = F_R\left(\frac{\partial U_i}{\partial r}, g_{\tilde\mu\tilde\nu}, \frac{\partial \Phi}{\partial \theta}, \frac{\partial \Theta}{\partial \theta}\right)\, .
\end{split}
\ee
Again, using the Cauchy-Kovalevskaya theorem on the last two equations completes our analysis.

Note that the above analysis is general and assumes no accidental symmetries of the seed metric $g_{\mu\nu}$. 
One may need to slightly adjust the derivation in case of extra symmetries.
For example, if one starts with $g_{\mu\nu}$ in the Boyer--Lindquist form (which implies circularity), \cref{eq:gphitheta,eq:grtheta} immediately imply $U=U(r)$ and $\Phi=\Phi(r)$, while the other two give equations similar to \eqref{eq:quadraticeqs}, though for $\pdv*{\Theta}{\theta}$ and $\pdv*{R}{\theta}$. 
The conclusion nevertheless holds --- one can find a local change of coordinates that brings the metric to the Kerr-like (or the orthogonal) form.

We would like to stress that the above analysis shows the existence of a coordinate transformation that locally brings the metric to the orthogonal gauge (with or without the condition $g_{rr}=0$), but it does not guarantee the possibility to find such a chart globally.
This can be clearly traced to the limitations of the Cauchy-Kovalevskaya theorem used above, which affirms only local existence of solutions of differential equations. 
In practice it may happen that one needs to find several charts of the orthogonal gauge that cover the part of spacetime that one is interested in.

Besides, although the existence of the local coordinate change is guaranteed, establishing the explicit (analytic) form of such change when starting from a particular metric could be challenging --- if not impossible.
Again, this can be related to the Cauchy-Kovalevskaya theorem, which guarantees the existence of the solution to the differential equations but is of little help in finding such solution explicitly.

\subsection{Residual gauge freedom}

An important question is whether the Kerr-like gauge corresponds to a complete gauge fixing, or if there exist non-trivial coordinate changes of the kind of \cref{eq:diffeo} that leave it unchanged.
It is immediate to see that the gauge fixing is by far not complete, which can be appreciated in various ways.

The simplest way to see this is to notice that in the above proof of the local existence of the coordinate transformation there is a freedom corresponding to a residual gauge.
Indeed, at each step of finding the unknown functions $V \left(r,\theta \right)$, $R \left(r,\theta \right)$, $\Theta \left(r,\theta \right)$ and $\Phi \left(r,\theta \right)$, \cref{eq:Phidiff,eq:Thetadiff,eq:VRdiff}, one has to specify one piece of \qu{initial} data on the Cauchy surface. 
Barring possible imaginary solutions that may appear for some branches when getting \eqref{eq:Phidiff}, the initial data is arbitrary. 
Each choice of the initial data corresponds to a different solution of  $V \left(r,\theta \right)$, $R \left(r,\theta \right)$, $\Theta \left(r,\theta \right)$ and $\Phi \left(r,\theta \right)$, leading to the same Kerr-like or orthogonal form but with different metric coefficients.
The difference between these solutions comprises the residual freedom.

One can also see the presence of the residual gauge freedom in a more direct manner. 
Indeed, suppose we start with the Kerr-like gauge and then we make the transformation \eqref{eq:diffeo}.
In order to show that there is a residual gauge, one simply needs to repeat the steps given in the previous \cref{subsec:Proof}, with \cref{eq:OrthoGauge} and $g_{rr}=0$, and confirm the existence of a solution for the functions $V \left(r,\theta \right)$, $R \left(r,\theta \right)$, $\Theta \left(r,\theta \right)$, and $\Phi \left(r,\theta \right)$.
It is not difficult to check that one arrives at a form of \cref{eq:Phidiff} with considerably simpler coefficients $a_2$, $b_2$ and $c_2$, but the conclusion remains the same. 
Setting \qu{initial} conditions and using the Cauchy-Kovalevskaya theorem one can always find a local solution for $\Phi(r,\theta)$. 
The rest is completely analogous to that of \cref{subsec:Proof}, i.e.~one then finds $V \left(r,\theta \right)$, $R \left(r,\theta \right)$, $\Theta \left(r,\theta \right)$. 
Noticing that for general initial data the solution is not trivial, this implies that there is a residual change of coordinates that does not change the form of \cref{eq:OrthoGauge} and keeps $g_{rr}=0$.

\subsection{\label{subsec:Properties}Properties of the metric in Kerr-like coordinates}

We have argued that the orthogonal gauge of \cref{eq:OrthoGauge} is completely general and does not entail any loss of arbitrariness.
However, it makes it possible to present several features of the metric in a detailed yet relatively simple form.
In this subsection, we characterise the spacetime by collecting some of such features: 
we start by describing some relevant surfaces;
then move on to the construction of a null tetrad;
and finally comment briefly on geodesics.


\subsubsection{\label{subsubsec:Surfaces}Notable surfaces: static limit, rotosurface, event horizon}

We order our list of relevant surface starting from the outermost, then move inwards.
The first surface we wish to mention is the familiar (outer) \emph{static limit}, also known as ergosphere, defined as the locus of points at which the Killing vector $\xi^\mu$ becomes null.
In our coordinates, it is located at $r=r_\text{erg}(\theta)$, where $r_\text{erg}$ is given by
\be\label{eq:Ergo}
\eval{g_{vv}}_{r=r_\text{erg}} = 0.
\ee
Inside the static limit, there can exist no static observers --- meaning, (timelike) observers whose velocity is proportional to $\xi^\mu$. 
For this reason this surface is also known as static limit.

However, locally non-rotating observers --- whose velocity is $\propto \xi^\mu + \Omega\, \psi^\mu$ with $\Omega:= -(\xi_\mu\, \psi^\mu) / (\psi_\nu\, \psi^\nu)$ --- are still allowed.
These observers cease to exist when $\xi^\mu + \Omega\, \psi^\mu$ becomes null.
Following \cite{CarterBlackHoles1973,CarterRepublicationBlack2010}, we refer to the surface where this happens as \emph{rotosurface}, although the term \emph{stationary limit} has also been used e.g.~in \cite{AnsonDisformingKerr2021}.
Since
\be
\left(\xi_\mu + \Omega\, \psi_\mu \right) \left(\xi^\mu + \Omega\, \psi^\mu \right) = \frac{1}{g_{\phi \phi}} \left[ g_{vv} g_{\phi \phi} - g_{v\phi}^2 \right] =: \frac{D_{(v\phi)}}{g_{\phi \phi}} \, ,
\ee
where we have introduced the symbol $D_{(v\phi)}$ to indicate the determinant of the $v$-$\phi$ sub-block of the metric, the rotosurface's location $r=R(\theta)$ is determined by
\be\label{eq:Roto}
\eval{D_{(v\phi)}}_{r=R} = 0\, .
\ee
Note that, since $\psi^\mu$ vanishes on the axis of symmetry, the rotosurface touches the static limit at the poles $\theta = 0$, $\pi$.
Further note that, within the orthogonal gauge,
\be\label{eq:grrAnsatz}
g^{rr} = \frac{g_{\theta\theta} D_{(v\phi)} }{\det(g_{\mu\nu})}\, ,
\ee
hence the rotosurface's location is equivalently given by the zeroes of $g^{rr}$.

If $R(\theta)$ happens to be independent on the angle $\theta$, i.e.~$\dv*{R}{\theta}= 0$, then the rotosurface is a null surface and it coincides with the black hole's event horizon. 
Indeed, the vector $\partial_\mu \left[ r - R(\theta) \right]$, which is normal to the rotosurface, has norm given, in the orthogonal gauge of~\cref{eq:OrthoGauge}, by
\be
g^{\mu \nu} \partial_\mu \left[ r - R(\theta) \right] \partial_\nu \left[ r - R(\theta) \right] = 
g^{rr} + g^{\theta\theta} \left(\dv{R}{\theta}\right)^2 \, ;
\ee
so, if $\dv*{R}{\theta}= 0$, the rotosurface is null, and therefore it coincides with the event horizon.
In this case, the (null) normal becomes proportional to the vector $\xi^\mu + \Omega \, \psi^\mu$, which is thus a generator of the rotosurface.
Hence, if additionally $\eval{\Omega}_R$ happens to be constant, then the generator $\xi^\mu + \Omega \, \psi^\mu$ is also a Killing vector and, consequently, the event horizon a Killing horizon. 
As mentioned, all these circumstances are necessarily realised if the spacetime is circular --- cf.~\cref{sec:Circ}.

In general, however, this needs not be the case:
the rotosurface might not be a null surface and therefore it might not coincide with the black hole's event horizon.
Assuming that the horizon's location is given by $r=H(\theta)$ for some $H$, the requirement that it be a null surface entails, within the orthogonal gauge,
\be\label{eq:H}
\eval{g^{rr}}_{r=H} + \eval{g^{\theta\theta} }_{r=H} \left(\dv{H}{\theta}\right)^2 = 0 \, ,
\ee
which is a differential equation for the function $H$.
For the horizon surface not to be singular at the poles we must have $\dv*{H}{\theta} = 0$ at $\theta=0$, $\pi$, and we can thus deduce that the horizon touches the rotosurface there.
Moreover, if the metric enjoys a reflection symmetry about the equator, i.e.~$g_{\mu \nu}(r,\theta) = g_{\mu \nu}(r,\pi-\theta)$, then $\dv*{H}{\theta}$ must vanish there too and the rotosurface touches the horizon on this plane as well.
The horizon --- if it exists --- is always located inside the rotosurface, as long as $g^{\theta \theta}>0$, since to satisfy \cref{eq:H} we need $g^{rr}\leq 0$. 
Because $g^{rr}=0$ at the rotosurface and $g^{rr}>0$ at infinity, we deduce $H \leq R$.
This means that there exists an untrapped region between $r=R$ and $r=H$ in which there is no timelike Killing vector and hence there can be no stationary observers.

\subsubsection{\label{subsubsec:Tetrad}Null tetrad}

So far, the role of the orthogonal gauge has been merely to simplify the writing of \cref{eq:grrAnsatz,eq:H}.
Assuming now the full Kerr-like gauge, \cref{eq:OrthoGauge} with $g_{rr}=0$, we can push our characterisation somewhat further, since the corresponding coordinates are expected to be naturally adapted to particular families of null directions.
For example, in the Kerr geometry, when the metric is written in Kerr coordinates, one can easily read off the two principal null directions and with them construct the so-called Kinnersley tetrad \cite{FrolovBlackHole1998}.
Null tetrads are particularly useful for various types of analyses, such as that of the algebraic properties of the spacetime, and this therefore motivates us to construct a generalised Kinnersley-like tetrad adapted to our case.

Namely, we wish to find two real null vectors $k^\mu$ and $l^\mu$, normalised so that $k_\mu \, l^\mu = 1$, and tangent to a family of, respectively, ingoing and outgoing null curves.
We anticipate that, contrary to their analogues in the Kerr spacetime, the congruences to which the vectors that we will find are tangent are not going to be geodesic, in general, nor need they be aligned with the spacetime's principal null directions.
Moreover, we seek two complex null vectors $m^\mu$ and $\overline{m}^\mu$, conjugate to one another, such that $m_\mu \, \overline{m}^\mu = -1$ and $m_\mu \, k^\mu = m_\mu \, l^\mu = 0$.

Identifying $k^\mu$ is rather immediate, as the condition $g_{rr}=0$ points directly to
\be\label{eq:k}
k^\mu  = \left\{0,\ q(r,\theta) ,\ 0,\ 0 \right\}^\mu \, ,
\quad
k_\mu = \left\{ q(r,\theta) g_{vr},\ 0,\ 0,\ q(r,\theta) g_{r\phi} \right\}_\mu\, ,
\ee
where $q$ is a function which we assume to be negative in order to ensure that the congruence is ingoing.
As anticipated, this $k^\mu$ is not geodesic, in general, since
\be
k^\alpha \nabla_\alpha k_\mu = q \left\{ \partial_r \left( q  g_{vr} \right)  ,\ 0,\ 0,\ \partial_r \left( q g_{r\phi}\right) \right\}_\mu\, .
\ee
If however
\be\label{eq:GeodCond}
\frac{\partial_r \left( q  g_{vr} \right)}{g_{vr}} = \frac{\partial_r \left( q g_{r\phi}\right)}{g_{r\phi}} 
\Rightarrow
\partial_r \left( \frac{g_{vr}}{g_{r\phi}} \right) = 0\, ,
\ee
then the acceleration above is proportional to $k_\mu$ itself, which is therefore tangent to a family of null geodesics, although these are not automatically affinely parametrised.
Affinely parametrised geodesics correspond to the choice 
\be
\partial_r (q g_{vr} ) = 0\, .
\ee
Note that, if the Kerr-like coordinates are ingoing, asymptotic flatness would suggest $g_{vr} \to +1$ as $r \to \infty$;
hence, the most obvious choice is
\be
q = - \frac{1}{g_{vr}} \, .
\ee
We wish to stress that \cref{eq:GeodCond} is a non-trivial condition on the metric components and in general it is not satisfied.

The second vector $l^\mu$ can also be found easily, although the result is admittedly less transparent.
First of all, note that the metric in the Kerr-like gauge --- \cref{eq:OrthoGauge} with $g_{rr}=0$ --- can be written in the following form:
\be\label{eq:KerrAnsatzSquare}
ds^2 = \left( g_{vr} \dd{v} + g_{r \phi } \dd{\phi} \right) \left[2 \dd{r} - g^{rr} \left( g_{vr} \dd{v} + g_{r \phi} \dd{\phi} \right) \right] - C_{\theta\theta} \left[ g^{r\phi} \dd{v} - g^{v r} \dd\phi \right]^2 
\, .
\ee
Here, we have introduced the symbol
\be
C_{\mu \nu} := \det(g_{ij})
\qq{with} i\neq \mu \, , j \neq \nu\, ,
\ee
which is, up to a sign, the cofactor of the metric element $g_{\mu\nu}$.
\Cref{eq:KerrAnsatzSquare} allows to immediately read off that a vector such that
\be
\frac{l^\phi}{l^v} 
= \frac{g^{r\phi}}{g^{vr}}
\qq{and}
\frac{l^r}{l^v} 
= \frac{1}{2} \frac{g^{rr}}{g^{vr}}
\ee
is null.
Imposing the normalisation with respect to $k_\mu$, we then find
\be\label{eq:l}
l^\mu = 
-\frac{1}{q} \left\{ 
g^{vr}, 
\frac{1}{2}g^{rr}, 
0, 
g^{r\phi}
\right\}^\mu \, .
\ee
One can easily verify that the acceleration of $l^\mu$ has a non-vanishing $\theta$ component and therefore $l^\mu$ cannot be geodesic.
However, when $k^\mu$ is geodesic, i.e.~when \cref{eq:GeodCond} is satisfied, then we can write
\be\label{eq:lquasigeo}
l^\nu \nabla_\nu l^\mu = \tilde{\kappa}\, l^\mu + A^\theta \delta^\mu_{\ \theta}\, ,
\ee
where $\tilde{\kappa}$ is a scalar function (not a constant).
That is, in this case $l^\mu$ is \qu{almost} geodesic.
Moreover, we were able to prove that the component $A^\theta$ vanishes when the circularity conditions of \cref{eq:CircDef} are satisfied.

Finally, the last two legs of the Kinnersley-like tetrad may be given by
\be\label{eq:m}
m^\mu = \frac{1}{\sqrt{2}} \left\{ - \frac{g_{r\phi} }{ \sqrt{-C_{\theta\theta} }} , 0 , - \frac{\iu }{\sqrt{ g_{\theta\theta}} } , \frac{g_{vr}}{\sqrt{-C_{\theta\theta} }} \right\} 
\ee
and its complex conjugate.
In writing \cref{eq:m}, we assumed $g_{\theta\theta} > 0$ and $C_{\theta \theta}<0$, as is the case for the Kerr spacetime.

One may easily verify that, specifying these expressions to the Kerr metric in (ingoing) Kerr coordinates, one finds that $k^\mu$ and $l^\mu$ give the usual Kinnersley legs, while $m^\mu$ and $\overline{m}^\mu$ only differ from their textbook counterparts \cite{FrolovBlackHole1998} by an irrelevant normalisation.

\subsubsection{\label{subsubsec:Geodesics}Geodesics}

We close our characterisation of the spacetime with a few comments concerning geodesics.
We consider a test particle of mass $\mu$, position $x^\mu$, and momentum $p_\mu$, subject to the Hamiltonian
\be\label{eq:Hamiltonian}
H \left( x^\mu,\, p_\mu \right) &= \frac{1}{2}g^{\mu \nu} p_\mu p_\nu \, .
\ee
In light of the spacetime symmetries, the Killing energy $E:= - p_\mu\, \xi^\mu$ and the Killing angular momentum $L := p_\mu\, \psi^\mu$ are conserved on shell.

Our first comment concerns the complete integrability of the motion --- or the lack thereof.
A dynamical system is said to be completely (Liouville) integrable if it possesses the maximum number of independent integrals of motion: namely, as many as its degrees of freedom \cite{FrolovBlackHoles2017}.
This property is tightly related to the separability, in the sense of partial differential equations, of the corresponding Hamilton--Jacobi equation
\be\label{eq:HJ1}
\pdv{S}{\lambda} = - H \left( x^\mu, \pdv{S}{x^\mu} \right) \, .
\ee
For this reason, we find it most convenient to cast the discussion in the language of the Hamilton--Jacobi formalism.
Thanks to the Killing symmetries, we may write the Hamilton's principal function $S$ as
\be
S = \frac{1}{2} \mu^2 \lambda - E v + L \phi + \tilde{S}(r,\theta)\, .
\ee
Accordingly, \cref{eq:HJ1} can be written as
\be\label{eq:HJ2}
g^{rr} \left\{ \left(\pdv{\tilde{S}}{r} \right)^2 + 2 \left[-E \frac{g^{vr}}{g^{rr}} + L \frac{g^{r\phi}}{g^{rr}} \right] \left(\pdv{\tilde{S}}{r} \right)\right\} + g^{\theta \theta}  \left(\pdv{\tilde{S}}{\theta} \right)^2 
- \frac{g_{vr}^2 g^{\theta\theta}}{ \det(g_{\mu \nu}) } \left[ \frac{g_{r\phi}}{g_{vr}} E +  L \right]^2
+ \mu ^2 = 0 \, .
\ee
By writing $\tilde{S}(r,\theta) = S_r (r) + S_\theta(\theta)$, it may be possible to separate the variables and reduce this partial differential equation to two decoupled ordinary differential equations.
This is the case if the system is completely integrable, i.e.~if there exists a fourth independent constant of motion;
at any rate, separability clearly requires that the metric components be related to one another through some very delicate relations, and is therefore quite a rare property.

In the Kerr spacetime, \cref{eq:HJ2} does allow for separation of variable.
This fact, which appears rather serendipitous, is related to the existence of a hidden constant of motion, discovered by Carter and associated with a rank-two Killing tensor \cite{CarterGlobalStructure1968}.
In spacetimes different from Kerr, this property is not guaranteed.
Inspecting \cref{eq:HJ2} very attentively, we realise that a set of sufficient conditions for separability is
\bse\label{eq:Separability}
\be
\label{eq:SeparabilityCirc}
\pdv{}{\theta} \left( \frac{g^{vr}}{g^{rr}}\right) &= 0 \, ,
\quad
\pdv{}{\theta} \left( \frac{g^{r\phi}}{g^{rr}}\right) = 0 \, ,\\
\label{eq:Separabilityk}
\pdv{}{r} \left( \frac{g_{r\phi}}{g_{vr}}\right) &= 0 \, ,\\
\frac{g_{vr}^2 g^{\theta\theta}}{\det(g_{\mu \nu})} &= \text{const.} \, ,
\quad
g^{\theta\theta} = \frac{1}{\sigma_r(r) + \sigma_\theta(\theta) } \, ,
\quad
g^{rr} = \frac{\rho(r)}{\sigma_r(r) + \sigma_\theta(\theta) } \, ,
\ee\ese
where $\rho$ and $\sigma_r$ are functions of $r$ only while $\sigma_\theta$ is a function of $\theta$ only.
Clearly, these conditions are sufficient but, most likely, they are not necessary, as there might exist other functional relations that also lead to a separable Hamilton--Jacobi equations.
Still, the conditions \eqref{eq:Separability} are those satisfied by the Kerr metric.

We point out that \cref{eq:Separabilityk} is the condition \eqref{eq:GeodCond} ensuring that the ingoing null congruence $k^\mu$ be geodesic.
Moreover, as we will show in \cref{sec:SolvingCirc}, \cref{eq:SeparabilityCirc} are equivalent to demanding that the metric be circular.
We may therefore infer that the metrics with completely integrable geodesics constitute a proper subset of circular metrics.
This conclusion is further supported by the following fact:
It is possible to write down the most general metric admitting a rank-two Killing tensor, analogous to the one discovered by Carter, and such metric is found to be circular \cite{BenentiRemarksCertain1979,JohannsenMetricRapidly2011,DelaporteParameterizationsBlackhole2022a}.
Therefore, a non-circular metric will generically not have completely integrable geodesics.

A second, partially unrelated comment concerns equatorial motion.
By computing the Hamilton's equations ensuing from \cref{eq:Hamiltonian}, or directly from the geodesics equations, one may derive the following:
\be
\ddot{\theta} = -\frac{1}{2} g^{\theta \theta} p_\alpha p_\beta \partial_\theta g^{\alpha \beta} + p_\theta p^\alpha \partial_\alpha g^{\theta \theta}\, .
\ee
If there exists a $\theta_0$ such that 
\be\label{eq:EqMotionCond}
\partial_\theta g_{\mu \nu} (r, \theta_0) = 0
\qq{for all $\mu$ and $\nu$}
\ee
and for all values of $r$, then on $\theta = \theta_0$ one has that $p_\theta = 0 \Rightarrow \ddot{\theta} = 0$, hence there exist orbits that lie entirely on the surface $\theta = \theta_0$. 
Such surface is generically a cone, save for two exceptions:
$\theta_0 = 0$ or $\theta_0 = \pi$, corresponding to the axis of symmetry;
and $\theta_0 = \pi/2$, for which the surface coincides with the equator.
For continuity reasons \cref{eq:EqMotionCond} is always satisfied on the axis of symmetry, and motion along said axis is therefore always possible --- although it requires $L=0$.
We thus deem it more interesting to focus on equatorial motion.

First of all, we wish to point out that purely equatorial motion is not always possible, as \cref{eq:EqMotionCond} is not guaranteed to hold.
One possibility to ensure that \cref{eq:EqMotionCond} is satisfied consists in invoking an additional symmetry under reflections about the equator, i.e.~that $g_{\mu \nu} (r, \theta) = g_{\mu \nu} (r, \pi - \theta)$.
Such symmetry is sensible, as it expresses the notion that one cannot distinguish the northern and southern hemispheres; 
however, it constitutes an additional assumption.
Further note that the existence of equatorial motion is unrelated to the circularity of the metric.

On the other hand, if equatorial motion is possible, then this particular motion is automatically integrable, in the sense that it is governed by the following \qu{quadrature} equation, which follows from the normalisation of the particle's momentum $p_\mu\, p^\mu = - \mu^2$:
\be
C_{\theta\theta} \left( \dot{r} \right)^2 = - \left[ g_{\phi \phi} E^2 + 2 g_{v \phi} E L + g_{v v} L^2 + \mu^2 D_{(v\phi)} \right]\, .
\ee
The right-hand side of this equation can be interpreted as an effective potential and, notably, the metric components by which it is determined are nothing but the scalar products of the Killing vectors among themselves: 
\be
g_{vv} = \xi_\mu \, \xi^\mu\, ,
\quad
g_{v\phi} = \psi_\mu \, \xi^\mu\, ,
\quad
g_{\phi \phi} = \psi_\mu \, \psi^\mu\, .
\ee
Hence equatorial motion appears but a very partial probe of the geometry, and not likely to allow for smoking-gun tests of non-Kerr behaviour.
For instance, one can easily envisage deformations of the Kerr geometry, not necessarily small, that leave equatorial orbits unaltered.
In particular, features often associated with observables, such as the location of the innermost stable circular orbit and of the light ring, or the value of the effective potential's second radial derivative evaluated at the potential's peak, carry virtually no information on circularity.

\section{\label{sec:SolvingCirc}Solving the circularity conditions}

The circularity conditions of \cref{eq:CircDef} represent partial differential equations for the components of the metric in a given coordinate chart.
Quite remarkably, these equations can be completely solved in a fully analytic manner.
Since the system is clearly underdetermined, such solution does not consist in an explicit expression for the components as functions of the coordinates, but rather in a pair of conditions that relate different components.
Crucially, these conditions are algebraic, not differential.
To the best of our knowledge, such solution has never appeared before, and for this reason we report the relevant computations in some detail.

We start by noticing that the left-hand sides of \cref{eq:CircDef} are four-forms and therefore, in four spacetime dimensions, they are proportional to the volume form.
Hence, \cref{eq:CircDef} can be reformulated as follows:
\bse\label{eq:CircEpsilon}
\be
\varepsilon^{\mu \nu \rho \sigma} \xi_\mu \psi_\nu \partial_\rho \xi_\sigma &=0 \, , \\
\varepsilon^{\mu \nu \rho \sigma} \xi_\mu \psi_\nu \partial_\rho \psi_\sigma &=0\, .
\ee\ese
In our coordinate system, which was chosen so that $\xi^\mu = \delta^\mu_{\ v}$ and $\psi^\mu = \delta^\mu_{\ \phi}$, the previous equations become 
\bse\be
\varepsilon_{v \phi \rho \sigma} g_{v\beta} g^{\rho \alpha}\partial_\alpha g^{\sigma\beta} &= 0
 \, ,\\
\varepsilon_{v \phi \rho \sigma} g_{\phi \beta} g^{\rho \alpha}\partial_\alpha g^{\sigma\beta} &= 0 \, .
\ee\ese
Because of the antisymmetry of the Levi-Civita tensor, the indeces $\rho$ and $\sigma$ effectively only take values in $\{r,\theta\}$;
moreover, the index $\alpha$ also takes values in $\{r,\theta\}$, as all other derivatives vanish.

These equations simplify even further invoking the orthogonal gauge of \cref{eq:OrthoGauge}.
Indeed, in this gauge most terms in the previous contractions vanish and the two equations reduce to
\bse\be
g^{\theta\theta} \left[ g_{v \beta}  \partial_\theta g^{\beta r} \right] &= 0 \, ,\\
g^{\theta\theta} \left[ g_{\phi \beta} \partial_\theta g^{\beta r} \right] &= 0 \, .
\ee\ese
This form represents a drastic simplification and one might wander whether one could arrive to an equivalent result without resorting to this particular gauge choice.
The answer is plausibly in the affirmative, although we have not carried out the necessary checks since, as argued in \cref{sec:Gauge}, there is no loss of generality in taking $g_{\mu \theta} \propto \delta^\theta_{\ \mu}$.
Note that we never needed to set $g_{rr}=0$, i.e.~to go to the full Kerr-like gauge.

Spelling out the last remaining contraction, and assuming $g^{\theta \theta} \neq 0$ (as it must, otherwise the metric would be singular), we get
\bse\be
g_{vv} \partial_\theta g^{vr} + g_{vr} \partial_\theta g^{rr} + g_{v\phi} \partial_\theta g^{r \phi} &= 0 \, ,\\
g_{v \phi} \partial_\theta g^{v r} + g_{r \phi } \partial_\theta g^{rr} + g_{\phi \phi} \partial_\theta g^{r \phi} &= 0\, .
\ee\ese
We can then manipulate these equations to eliminate, say, $\partial_\theta g^{rr}$ from one of them and $\partial_\theta g^{r\phi}$ from the other, thereby obtaining
\bse\label{eq:CircPDE1}\be
\left[ g_{r\phi} g_{vv} - g_{vr} g_{v \phi} \right] \partial_\theta g^{vr} + \left[ g_{r\phi} g_{v\phi} - g_{vr} g_{\phi \phi} \right] \partial_\theta g^{r\phi} &= 0 \, , \\
\left[ g_{\phi\phi} g_{vv} - \left( g_{v \phi} \right)^2 \right] \partial_\theta g^{vr} + \left[ g_{\phi\phi} g_{v r} - g_{v\phi} g_{r\phi} \right] \partial_\theta g^{rr} &= 0 \, .
\ee\ese
At this point, we notice that (in the orthogonal gauge)
\be
\frac{ g_{r\phi} g_{vv} - g_{vr} g_{v \phi} }{g_{r\phi} g_{v\phi} - g_{vr} g_{\phi \phi}} = - \frac{g^{r\phi}}{g^{vr}}
\qq{and}
\frac{ g_{\phi\phi} g_{vv} - \left( g_{v \phi} \right)^2 }{ g_{\phi\phi} g_{v r} - g_{v\phi} g_{r\phi}}  = - \frac{g^{rr}}{g^{vr}}
\ee
hence \cref{eq:CircPDE1} are equivalent to the system
\bse\be
g^{r\phi} \partial_\theta g^{vr} - g^{v r} \partial_\theta g^{r \phi} &= 0\, , \\
g^{rr} \partial_\theta g^{vr} - g^{v r} \partial_\theta g^{rr} &= 0 \, ,
\ee\ese
which can be integrated straightforwardly.

The solution to this system may be written as
\be\label{eq:CircSol}
g^{vr} = g^{rr} f(r) \qq{and} g^{r\phi} = g^{rr} h(r)
\ee
with $f$ and $h$ arbitrary functions of $r$.
It is a necessary and sufficient condition for a metric (expressed in the orthogonal gauge) to be circular.
It is a \qu{solution} to the (differential) circularity conditions \eqref{eq:CircDef} in the sense that, as mentioned, it consists in algebraic relations among the metric components.

\Cref{eq:CircSol} represents one of the key results of this article, one that we will exploit extensively in \cref{sec:Examples} to construct simple examples of non-circular metrics in which circularity is broken \qu{softly}.
Before moving on, however, a few remarks are in order.

First of all, an important consistency check is to make sure that the equivalent formulation of the circularity conditions in terms of the Ricci tensor, \cref{eq:CircDefRicci}, yields the same result as above.
With our choice of coordinate, \cref{eq:CircDefRicci} become
\bse\be
R^r_{\ v} = 0 = R^\theta_{\ v} \, , \\
R^r_{\ \phi} = 0 = R^\theta_{\ \phi} \, .
\ee\ese
These are four equations, but one may verify that only two of them are independent.
We have computed the Ricci tensor and confirmed that these equations do not lead to additional constraints with respect to the ones just described.
We omit the details, as they are not particularly insightful.

A second remark is the following:
The solution we have just found has a neat expression when written in terms of the components of the inverse metric, as we did in \cref{eq:CircSol}.
Rewriting these conditions in terms of the components of the direct metric yields
\be\label{eq:CircSolDown}
\frac{g_{vr} g_{\phi \phi} - g_{r\phi} g_{v \phi}}{g_{v \phi}^2 - g_{vv} g_{\phi \phi}} = f(r)
\qq{and}
\frac{g_{r\phi} g_{vv} - g_{vr} g_{v \phi}}{g_{v \phi}^2 - g_{vv} g_{\phi \phi}} = h(r) \, ,
\ee
which are much more involved (and not linear). 
One can see that the choice to work in terms of the inverse metric was crucial to find the conditions of \cref{eq:CircSol} or \cref{eq:CircSolDown}.

Yet another, possibly more interesting comment concerns the interpretation of the solution of \cref{eq:CircSol}.
As mentioned in \cref{sec:Circ}, circularity is naturally associated with the existence of an adapted coordinate system in which the metric has the \qu{minimum} number of non-diagonal components --- what we called the Boyer--Lindquist form.
\Cref{eq:CircSol} renders this property particularly manifest.
Indeed, consider a change of coordinates of the kind considered in \cref{sec:Gauge} with
\be
V = V(r)\, , \quad
\Phi = \Phi(r)\, , \quad
R = r\, , \quad
\Theta = \theta \, ,
\ee
and ask that the resulting metric be in Boyer--Lindquist form.
Starting from the orthogonal gauge of \cref{eq:OrthoGauge}, we only need to impose the additional conditions 
\be
g^{vr} = 0 
\qq{and}
g^{r\phi} = 0 \, ,
\ee
or equivalently
\be
g_{vr} = 0 
\qq{and}
g_{r\phi} = 0 \, .
\ee
Solving for $\dv*{V}{r}$ and $\dv*{\Phi}{r}$, we get
\be
\dv{V}{r} = \frac{g^{\tilde{v} \tilde{r} }}{g^{ \tilde{r} \tilde{r}}}
\qq{and}
\dv{\Phi}{r} = \frac{g^{ \tilde{r} \tilde{\phi}}}{g^{ \tilde{r} \tilde{r} }} \, ,
\ee
or equivalently
\be
\dv{V}{r} = \frac{g_{ \tilde{v} \tilde{r} } g_{ \tilde{\phi} \tilde{\phi} } - g_{ \tilde{r} \tilde{\phi}} g_{ \tilde{v} \tilde{\phi} }}{g_{ \tilde{v} \tilde{\phi} }^2 - g_{ \tilde{v} \tilde{v} } g_{ \tilde{\phi} \tilde{\phi} }}
\qq{and}
\dv{\Phi}{r} = \frac{g_{ \tilde{r} \tilde{\phi} } g_{ \tilde{v} \tilde{v} } - g_{ \tilde{v} \tilde{r} } g_{ \tilde{v} \tilde{\phi} }}{g_{ \tilde{v} \tilde{\phi} }^2 - g_{\tilde{v}\tilde{v}} g_{\tilde{\phi} \tilde{\phi}}} \, .
\ee
These equations are integrable provided their right-hand sides depend on $r$ only (and not on $\theta$).
Confronting with \cref{eq:CircSol,eq:CircSolDown}, we thus realise that the existence of a coordinate transformation that brings the metric to a Boyer--Lindquist form is equivalent to the requirement that the metric be circular.

\section{\label{sec:Examples} Examples}

The great advantage of solving the (differential) circularity conditions of \cref{eq:CircDef} in terms of the algebraic relations of \cref{eq:CircSol} is that one has now a much greater control on the breaking of circularity.
For instance, it becomes possible to construct spacetimes that are \qu{perturbatively not circular}, in the sense that the deviations away from a reference circular spacetime are governed by a set of tunable parameters.
Alternatively, one can construct simple explicit examples with the aim of investigating specific consequences of circularity breaking.

Here, we choose to focus on the properties of the black hole's event horizon, and present two simple examples that help us illustrate how these are impacted by the breaking of circularity.
Both examples can be seen as deformations of the Kerr metric.
The first one is truly \qu{minimal}, in some sense, since the location of \emph{all} the relevant surfaces presented in \cref{subsubsec:Surfaces} (static limit, rotosurface, and event horizon) is unchanged with respect to Kerr, and the key consequence on circularity breaking is that the horizon is no longer Killing.
The second example is somewhat less \qu{minimal}, since these surfaces are purposely displaced from their Kerr analogues, though in a way that allows to maintain analytical control over computations --- this is at variance with the metrics analysed in e.g.~\cite{AnsonDeformedBlack2021,eichhorn_image_2021}, for which the horizon's location had to be determined numerically.

We point out that the Kerr metric, written in the ingoing Kerr coordinates representing the analogue of our Kerr-like gauge, reads
\be\label{eq:Kerr}
\dd{s}^2 &= - \left( 1 - \frac{2Mr}{\Sigma} \right) \dd{v}^2 
+ \Sigma \dd{\theta}^2
+ \frac{A \sin^2\theta}{\Sigma} \dd{\phi}^2 \nonumber\\
&\phantom{=} + 2 \dd{v}\dd{r}
- 2 a\sin^2\theta \dd{r}\dd{\phi}
- \frac{4 aM r\sin^2\theta}{\Sigma} \dd{v} \dd{\phi} 
\, ,
\ee
with $\Delta:= r^2 -2Mr +a^2$, $\Sigma := r^2 +a^2\cos^2\theta$, and $A := (r^2+a^2)^2 - \Delta a^2 \sin^2\theta$, while $M$ and $a$ are respectively the mass and spin parameter.
Hence, in particular
\be
g^{rr}_\text{Kerr} = \frac{\Delta}{\Sigma}\, , \quad g^{vr}_\text{Kerr} = \frac{r^2+a^2}{\Sigma} \, , \quad g^{r \phi}_\text{Kerr} = \frac{a}{\Sigma}\, ,
\ee
and the spacetime is obviously circular, since
\be
\frac{g^{vr}_\text{Kerr}}{g^{rr}_\text{Kerr}} = f_\text{Kerr} = \frac{r^2+a^2}{\Delta} 
\qq{and} 
\frac{g^{r \phi}_\text{Kerr}}{g^{rr}_\text{Kerr}} = h_\text{Kerr} = \frac{a}{\Delta}
\ee
are functions of $r$ only --- cf.~\cref{eq:CircSol}.

\subsection{\label{subsec:EgMinimal}\qu{Minimal} breaking of circularity}

In light of what has been said above, a very simple way to deform the Kerr solution into a non-circular spacetime consists in replacing $f_\text{Kerr}(r)$ with some $f(r,\theta)$, and/or $h_\text{Kerr}(r)$ with $h(r,\theta)$.
Crucially, such replacement must introduce a $\theta$ dependence in order for circularity to be broken.

We choose to focus on the following example:
We fix all components of the inverse metric as in Kerr, except for
\be
g^{vr} = g^{vr}_\text{Kerr} + \frac{\delta(r,\theta)}{\Sigma} = \frac{\Delta + \delta(r,\theta)}{\Sigma} \, ;
\ee
hence $\delta(r,\theta)$ parametrises the deviations away from circularity, and $\delta = 0$ corresponds to Kerr.
Explicitly, the inverse and direct metric are:
\bse\label{eq:MinimlaDefo}
\be
\label{eq:MinimlaDefoInv}
g^{\mu \nu}\partial_\mu \partial_\nu &=  
\frac{ a^2 \sin^2 \theta}{\Sigma} \partial_v \partial_v 
+ 2\frac{ a }{\Sigma} \partial_v \partial_\phi
+ \frac{1}{\sin^2\theta \Sigma} \partial_\phi \partial_\phi \nonumber\\
&\phantom{=}+ \frac{\Delta}{\Sigma}\partial_r \partial_r 
+ \frac{1}{\Sigma} \partial_\theta \partial_\theta 
+ 2\frac{a}{\Sigma} \partial_r \partial_\phi 
+ 2  \frac{r^2+a^2+\delta }{\Sigma} \partial_v \partial_r \, ,\\
\label{eq:MinimlaDefoDir}
g_{\mu \nu} \dd{x^\mu} \dd{x^\nu} &=  
\left(\frac{\Sigma}{\Sigma + \delta} \right)^2 \left[
- \left(1-\frac{2Mr}{\Sigma} \right) \dd{v^2} 
+ \Sigma \left(\frac{ \Sigma + \delta }{\Sigma} \right)^2 \dd{\theta^2}
+  \frac{A + \delta^2 + 2(r^2 + a^2) \delta }{\Sigma} \sin^2\theta \dd{\phi^2}
\right. \nonumber\\
&\phantom{=} \left. + 2 \left( \frac{\Sigma +\delta}{\Sigma} \right) \dd{v}\dd{r}
- 2a \sin^2\theta \left( \frac{\Sigma +\delta}{\Sigma} \right) \dd{r} \dd{\phi} 
- 2 a \sin^2\theta \left( \frac{2Mr + \delta}{\Sigma} \right) \dd{v} \dd{\phi} 
\right] \, . 
\ee\ese

The choice of the function $\delta$ is essentially arbitrary, although adhering to a logic of minimality imposes some mild constraints on it.
First of all, since the metric determinant is
\be
\det(g_{\mu \nu}) = - \left( \frac{\Sigma}{\Sigma + \delta} \right)^2 \Sigma^2 \sin^2\theta \, ,
\ee
we assume $\Sigma + \delta >0$ to prevent degeneracies.
Secondly, we want to ensure that the metric remain asymptotically flat and, possibly, that the Arnowitt--Deser--Misner (ADM) mass and angular momentum be given by $M$ and $a M$, respectively.
To this end, we must require $\delta/r \to 0$  as $r\to \infty$.
Moreover, we still want that the limit of vanishing spin correspond to a spherically symmetric spacetime, which requires $\partial_\theta \delta \to 0$ as $a \to 0$.
Finally, it is desirable that the limit $M \to 0$ still yield Minkowski spacetime. 
A further natural but entirely optional requirement is that the metric be symmetric under reflections about the equatorial plane, which can easily be implemented by requiring that $\delta$ be an even function of $\theta-\pi/2$.
A simple functional form that satisfies all these constraints is
\be\label{eq:deltaEg}
\delta = \epsilon \frac{(a M)^2}{\Sigma} \cos^2\theta
\, ,
\ee
with $\epsilon$ a dimensionless parameter that one may think small.

While the inverse metric \eqref{eq:MinimlaDefoInv} seems relatively simple, at first sight, the direct metric \eqref{eq:MinimlaDefoDir} most definitely does not.
Nonetheless, as mentioned, this deformation is quite \qu{minimal}, in the sense that the relevant surfaces of \cref{subsubsec:Surfaces} are not displaced with respect to their Kerr analogues.
In particular, the condition yielding the location of the static limit, $\eval{g_{vv}}_{r= r_\text{erg}}=0$, is still formally the same as in Kerr, namely
\be
r_\text{erg}^2 + a^2 \cos^2\theta - 2M r_\text{erg} = 0 
\Rightarrow
r_\text{erg} = M + \sqrt{M^2 - a^2 \cos^2 \theta} \, .
\ee
Similarly, the condition determining the location of the rotosurface, $\eval{g^{rr}}_{r=R}$, is still given by
\be
\eval{\Delta}_{r=R} = 0 
\Rightarrow
R = M + \sqrt{M^2 - a^2}\, .
\ee
Moreover, the rotosurface is a constant-$r$ sphere and, as argued in \cref{subsubsec:Surfaces}, this means it is also a null hypersurface.
The rotosurface therefore coincides with the black hole's event horizon, $H(\theta) = R(\theta)$.

However, since the spacetime is not circular, the horizon is not guaranteed to be Killing.
A quick computation yields the following angular velocity of frame dragging, $\Omega := - (\xi_\mu\, \psi^\mu)/(\psi_\nu \, \psi^\nu)$:
\be
\Omega &= \frac{a \left( 2M r + \delta \right)}{A + \delta^2 + 2(r^2 + a^2) \delta} \, ; 
\ee
on the horizon, this evaluates to
\be
\eval{\Omega}_H = \eval{\frac{a}{2 M r + \delta} }_H\, ,
\ee 
which is clearly not constant unless $\eval{\delta}_H$ is.
Therefore the vector $\xi^\mu + \Omega\, \psi^\mu$, which generates the horizon, is generically not Killing.

The null tetrad introduced in \cref{subsubsec:Tetrad} is --- cf.~\cref{eq:k,eq:l,eq:m} ---
\bse\label{eq:TetrMin}
\be
\label{eq:kMin}
k^\mu &= \left\{ 0,\, - \frac{\Sigma+\delta}{\Sigma}, \, 0,\, 0 \right\}^\mu \\
\label{eq:lMin}
l^\mu &=  \frac{\Sigma}{\Sigma+\delta} \left\{ r^2 +a^2 + \delta,\, \Delta, \, 0,\, a \right\}^\mu \\
\label{eq:mMin}
m^\mu &= \frac{ 1 }{\sqrt{2}\sqrt{\Sigma} } \left\{ a \sin\theta ,\, 0,\,  \iu , \frac{1}{\sin\theta} \right\}^\mu\, .
\ee\ese
First of all note that, as hinted to in \cref{subsubsec:Tetrad}, \cref{eq:mMin} does not immediately coincide with the textbook result of e.g.~\cite{FrolovBlackHole1998}, which however can be obtained via the simple rescaling
\be
\iu \sqrt{\frac{r - \iu a \cos \theta}{r + \iu \cos\theta}} m^\mu \, .
\ee
More interestingly, we realise that in our example the condition of \cref{eq:GeodCond} is satisfied;
therefore $k^\mu$ is geodesic and $l^\mu$ is \qu{almost} geodesic, since
\be
\label{eq:lAccMin}
l^\mu \nabla_\mu l_\nu &= \tilde{\kappa}\, l_\nu - \frac{\Delta}{2(\Sigma+\delta)^3} \pdv{\delta(r,\theta)}{\theta} \delta^\theta_{\ \nu} \, ,\\
\label{eq:kappaMin}
\qq{with}
\tilde{\kappa} &= \frac{1}{2} \left[ \pdv{}{r}\left( \frac{\Delta}{ \Sigma + \delta }\right) + \frac{\Delta}{\Sigma} \pdv{}{r} \left( \frac{\Sigma}{\Sigma+\delta}\right) \right] \, .
\ee
From the previous expression, it is clear that $l^\mu$ is automatically geodesic if the spacetime is circular, i.e.~$\partial_\theta \delta = 0$.

The tetrad of \cref{eq:TetrMin} allows to gain some further insight into the properties of the horizon. 
First of all, note that 
\be
\eval{l^\mu}_H &= \eval{\frac{2Mr +\delta}{\Sigma + \delta} \left[ \xi^\mu + \Omega\, \psi^\mu \right]}_H \\
\eval{l_\mu}_H &\propto \partial_\mu(r-H)\, ;
\ee
hence, $l^\mu$ is proportional to the \qu{would-be Killing} vector that generates the horizon, and to the vector normal to the horizon.
Moreover, \cref{eq:lAccMin} entails that $l^\mu$ is geodesic on the horizon.
The vector $l^\mu$ is therefore a generator of the horizon and the quantity $\tilde{\kappa}$ in \cref{eq:kappaMin} is thus related to the horizon's surface gravity --- more precisely (presumably) the \emph{inaffinity surface gravity} as per the classification of~\cite{CroppSurfaceGravities2013}.
Evaluating \cref{eq:kappaMin} on the horizon, we get
\be
\tilde{\kappa}_H := \eval{\frac{r-M}{\Sigma+\delta}}_{r=H}\, .
\ee
The familiar (inaffinity) surface gravity is instead defined as the amount by which $\xi^\mu + \Omega\, \psi^\mu$ fails to be affinely parametrised, and it can therefore be obtained simply by rescaling $\tilde{\kappa}_H$:
\be
\kappa_H &:= \eval{ \frac{\Sigma+\delta}{2Mr +\delta} }_H \tilde{\kappa}_H \nonumber\\
&= \eval{\frac{r-M}{2Mr + \delta}}_H\, .
\ee
The limit $\delta \to 0$ yields the Kerr result, as it must; for $\delta \neq 0$, the surface gravity inherits a dependence on $\theta$ and is therefore not constant on the horizon.

It is worth pointing out that, as it happens for the Kerr spacetime, the limit $a \to M$ corresponds to the roots of $\Delta$ becoming degenerate, and to $\kappa_H \to 0$.
This is therefore an extremal limit.
However, circularity is not automatically restored, unless the angular dependence of $\delta$ disappears as $a \to M$.
Moreover, the angular velocity of frame dragging remains not constant over the horizon.
The metric of \cref{eq:MinimlaDefoDir}, with $a=M$, thus provides a concrete realisation of a non-circular extremal black hole.

Further note that this spacetime presumably contains a singularity at $\Sigma = 0$.
We have checked that this is the case for the example of \cref{eq:deltaEg}, as the Ricci scalar is found to diverge, and it does not seem possible to remove the singularity with a suitable alternative choice of $\delta$;
however, the issue would definitely require further investigation.
Since for $a>M$ no horizon exists, the super-critical regime most likely corresponds to a naked singularity. 

In closing, we point out that, since we assumed an equatorial reflection symmetry, there exist orbits that lie entirely on the equator.
Such orbits are governed by the effective one-dimensional problem
\be
r^4 \left( \dot{r} \right)^2 = r \mathcal{R}_\text{Kerr} (r)
+ 2 E \left[ E \left( r^2 + a^2 \right) - a L \right] \delta(r,\pi/2) + E^2 \delta^2(r,\pi/2) \, ,
\ee
where 
\be
\mathcal{R}_\text{Kerr}(r) := E^2 \left( r^3 + a^2 r + 2 a^2 M \right) - 4 a M E L - \left( r - 2M \right) L^2 - m^2 r \Delta
\ee
is the effective potential that one finds for the Kerr metric.
Note that if $\delta(r,\pi/2) = 0$, as in the example of \cref{eq:deltaEg}, the effective potential coincides with that of Kerr.
Therefore, all the features of equatorial geodesics --- such as the location of the innermost stable circular orbit and the light ring --- are completely indistinguishable from those of Kerr.

\subsection{\label{subsec:EgNonMinimal}\qu{Not-so-minimal} breaking of circularity}

A second, very neat way of deforming the Kerr metric into a non-circular spacetime consists in replacing the mass parameter $M$ with a mass function $m(r,\theta)$.
Explicitly:
\bse\label{eq:NonMinimlaDefo}
\be\label{eq:NonMinimlaDefoInv}
g^{\mu \nu}\partial_\mu \partial_\nu &=  
\frac{ a^2 \sin^2 \theta}{\Sigma} \partial_v \partial_v 
+ 2\frac{ a }{\Sigma} \partial_v \partial_\phi 
+ \frac{1}{\sin^2\theta \Sigma} \partial_\phi \partial_\phi \nonumber\\
&\phantom{=}+ \frac{\tilde \Delta}{\Sigma}\partial_r \partial_r 
+ \frac{1}{\Sigma} \partial_\theta \partial_\theta 
+ 2\frac{a}{\Sigma} \partial_r \partial_\phi 
+ 2  \frac{r^2+a^2 }{\Sigma} \partial_v \partial_r \, ,\\
\label{eq:NonMinimlaDefoDir}
g_{\mu \nu} \dd{x^\mu} \dd{x^\nu} &=  
- \left(1-\frac{2r m(r,\theta) }{\Sigma} \right) \dd{v^2} 
+ \Sigma^2 \dd{\theta^2} 
+  \frac{\tilde A \sin^2\theta }{\Sigma} \dd{\phi^2}
\nonumber\\
&\phantom{=} + 2 \dd{v}\dd{r}
- 2a \sin^2\theta \dd{r} \dd{\phi} 
- \frac{4 r m(r,\theta)  a \sin^2\theta }{\Sigma} \dd{v} \dd{\phi} 
\, , 
\ee\ese
where $\tilde{\Delta} := r^2 + a^2 - 2 r m(r,\theta)$ and $\tilde{A} := (r^2 + a^2)^2 - \tilde{\Delta} a^2 \sin^2 \theta$.
Comparing with \cref{eq:CircSol}, we realise that the condition for the breaking of circularity is that $m(r,\theta)$ depend non-trivially on the angle $\theta$.

Except for this requirement, the choice of the mass function is largely arbitrary.
As in the previous subsection, we let ourselves be guided by a spirit of minimality.
Hence, we ask that $m(r,\theta) = M + \order{r^{-1}}$ as $ r \to \infty$ in order to preserve the asymptotic properties of the metric;
moreover, we require $\partial_\theta m(r,\theta) \to 0$ as $a \to 0$.
In passing, we point out that trimming the small-$r$ behaviour of $m(r,\theta)$ can \qu{regularise} the singularity present in the Kerr spacetime at $\Sigma = 0$.
More precisely, if $m(r,\theta) = \order{r^3}$ as $r \to 0$, then all scalar polynomials constructed out of the Riemann tensor and the metric are bounded in that limit --- see e.g.~\cite{BambiRotatingRegular2013,TorresRegularRotating2023,MazzaHeartDarkness2023} and references therein.
The metric of \cref{eq:NonMinimlaDefo} may therefore describe a regular black hole.

Irrespective of the particular form of $m(r,\theta)$, the metric of \cref{eq:NonMinimlaDefo} appears very simple, arguably simpler than the one considered in the previous subsection.
However, the impact of this deformation on the near-horizon structure is typically rather drastic, and for this reason we deem this example \qu{less minimal} than the one of \cref{subsec:EgMinimal}.
Namely, the angular dependence of the mass function is generically inherited by the rotosurface and by the event horizon --- which are thus shifted with respect to their Kerr counterparts, and do not coincide with one another.

For a given $m(r,\theta)$ the event horizon's location can be determined by solving \cref{eq:H} as a differential equation for the horizon's profile $H(\theta)$.
Typically, e.g.~in \cite{AnsonDisformingKerr2021,eichhorn_locality-principle_2021,eichhorn_image_2021,FernandesRotatingBlack2023}, the integration has to be performed numerically.
Here, instead, we aim at constructing an example that can be treated analytically.
We shall proceed by reversing the problem:
we assume the horizon's profile $H(\theta)$ is given, and reverse-engineer a mass function $m(r,\theta)$ such that \cref{eq:H} is satisfied.
The choice of $H(\theta)$ is, once again, essentially arbitrary, except for the continuity requirement $H'(0) = 0 = H'(\pi)$.
As mentioned several times, we might further assume a reflection symmetry about the equator $H(\theta) = H(\pi - \theta)$, in which case it becomes fairly natural to express $H$ as a Fourier series in the basis $\{ \cos \left( 2n\theta \right) \}$ with $n=0, \dots, \infty$.
For illustration purposes, in the following we will sometimes refer to a particular example:
\be\label{eq:HEg}
H(\theta) = H_\text{Kerr} \left[ 1 - \epsilon \cos(2\theta) \right]\, ,
\ee
where $H_\text{Kerr} := M + \sqrt{M^2 - a^2}$ is the Kerr horizon's location and $\epsilon$ is a dimensionless parameter.

In any case, if $H(\theta)$ is known and the metric is that of \cref{eq:NonMinimlaDefo}, the horizon's equation \eqref{eq:H} reads
\be
\left( H' \right)^2 + \eval{\tilde{\Delta}}_H = 0\, ,
\ee
where the prime denotes the derivative with respect to $\theta$;
this is equivalent to
\be
m \left( H(\theta), \theta \right) = \frac{H^2 + a^2 + \left(H' \right)^2}{ 2 H } \, ,
\ee
which fixes the mass function on the horizon.
This then needs to be matched to the desired asymptotic form, i.e.~$m(r,\theta) = M + \order{r^{-1}}$ as $r \to \infty$ for an asymptotically flat spacetime.
We thus write
\be\label{eq:mMatching}
m(r,\theta) = M + \left[ \frac{H^2 + a^2 + \left(H' \right)^2}{ 2 H } - M\right] \mu(r,\theta)\, ,
\ee
with the understanding that 
\be
\mu(r,\theta) = \order{r^{-1}} 
\qq{as} 
r \to \infty\, ,
\qq{and}
\mu \left(H(\theta),\theta \right) = 1 \, .
\ee
For example, we may choose
\be\label{eq:muChoice}
\mu(r, \theta) = \frac{H(\theta)}{r} 
\ee
(or $\mu = (H/r)^\alpha$ with $\alpha>1$ if one wishes a more rapid fall off).

\begin{figure}[t]
    \centering
    \includegraphics[width=0.85\linewidth]{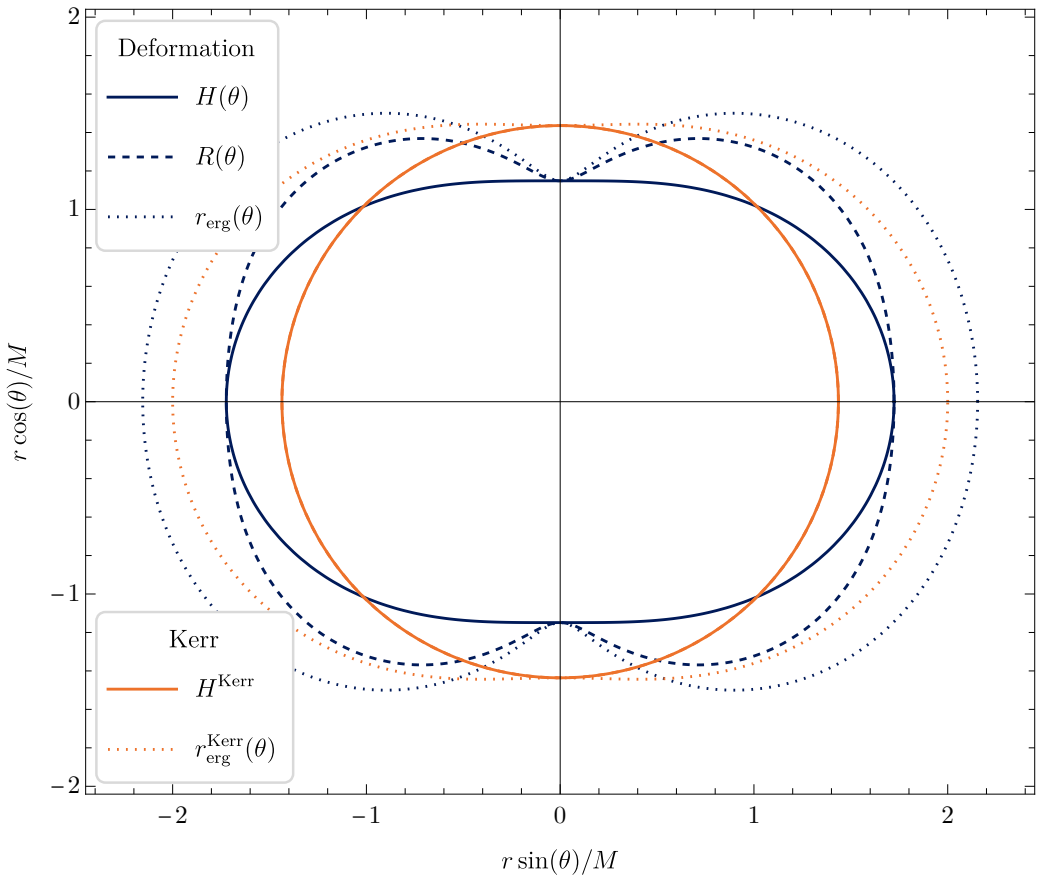}
    \caption{Horizon, rotosurface, and static limit for the specific \qu{non-so-minimal} deformation determined by \cref{eq:HEg,eq:muChoice}. Here, $a=0.9M$ and $\epsilon = 0.2$; for comparison, we show the equivalent profiles for a Kerr spacetime with the same spin.}
    \label{fig:HRErg}
\end{figure}

Once the mass function has been fixed, one can then solve \cref{eq:Roto} and \cref{eq:Ergo} to find, respectively, the location of the rotosurface and of the static limit.
These are algebraic equations, but depending on the choice of $\mu$ their roots might not be expressible in closed form.
For the choice of \cref{eq:muChoice}, we find two rather simple expressions:
\be
R &= M + \sqrt{(H-M)^2 + (H')^2} \, ,\\
r_\text{erg} &= M + \sqrt{(H-M)^2 + (H')^2 + a^2 \sin^2\theta} \, .
\ee
(Note that \cref{eq:Roto,eq:Ergo} have multiple roots: here we are focusing on the largest real ones, i.e.~on the outer rotosurface and static limit, but this example also yields an inner rotosurface and an inner static limit.)
These expressions, applied to the example of \cref{eq:HEg}, give the profiles depicted in \cref{fig:HRErg} for a representative choice of the parameters.

\section{Conclusions}

In this article, we investigated properties of generic stationary and axially symmetric spacetimes --- with a particular emphasis on circularity, and the breaking thereof.
Such focus is motivated by the fact that, as we have argued, renouncing circularity seems a necessary complication when searching for rotating solutions in alternative theories of gravity, and when constructing phenomenological models of rotating compact objects beyond GR.

First of all, within a set of coordinates adapted to the Killing symmetries, in \cref{sec:Gauge} we have proved the existence of two gauge choices: 
the orthogonal gauge of \cref{eq:OrthoGauge}, characterised by seven free functions of two variables; 
and a restriction thereof, which we called Kerr-like gauge, specified by the additional condition $g_{rr} = 0$ and thus characterised by six free functions.
Our argument consists in showing that there exists a coordinate transformation that brings any arbitrary stationary and axisymmetric metric into the Kerr-like (hence also orthogonal) form, and the result thus relies heavily on the Cauchy--Kovalevskaya theorem on the existence of a solution to an initial-value problem.
Our proof is therefore inherently local, because existence theorems for partial differential equations are almost always local, and because any statement concerning coordinates is essentially limited to the domain of existence of the particular coordinate patch one is working in.
Nonetheless, our proof suffices in demonstrating that both the orthogonal and Kerr-like gauges can be taken without loss of generality.

The key result of this article is the explicit solution, within the orthogonal gauge, of the circularity conditions \eqref{eq:CircDef} --- see \cref{sec:SolvingCirc}.
Such solution consists in the two algebraic relations of \cref{eq:CircSol}, which connect the three inverse metric components $g^{vr}$, $g^{r\phi}$, and $g^{rr}$.
These relations are particularly relevant, as they allow for the construction of \emph{ad hoc} examples in which different consequences of the loss of circularity can be disentangled and analysed in great detail. 

For illustrative purposes, in \cref{sec:Examples} we have constructed two such examples, both of which can be seen as deformations of the Kerr black hole and are conceived with the aim of putting the accent on the event horizon's properties.
The first example is \qu{minimal}, in the sense that the black hole's event horizon (as well as its static limit) is not displaced with respect to the one of Kerr, and the crucial consequence of circularity breaking is that the horizon is no longer Killing, so that the resulting surface gravity is not constant.
The second example is \qu{not so minimal}, in the sense that the horizon's profile can be chosen freely and is therefore different from that of Kerr; 
moreover, the horizon does not coincide with the rotosurface, and one can thus appreciate the difference between these two surfaces.

Clearly, our discussion of these two examples has been illustrative in scope and leaves ample room for further analyses, which we defer to future works.
For instance, the non-Killing nature of the horizon in the \qu{minimal} example raises the important question of thermodynamics, since different notions of surface gravity need not agree \cite{CroppSurfaceGravities2013} and formulating a zeroth law might be impossible.
Intuitively, one might expect that a non-constant surface gravity be associated to the emission of Hawking radiation with an angle-dependant temperature, which would suggest an interpretation of non-circular black holes as out-of-equilibrium thermodynamic systems.
However, careful analysis is needed in order to establish whether this intuition is borne out.

More generally, the conditions \eqref{eq:CircSol} open the way to further phenomenological investigations aimed at identifying proper observational signatures of (non-)circularity.
We have mentioned that circularity loss is generally related to the non-separability of the geodesic equations, which certainly increases the technical complexity of computations involving test particles and fields, as well as perturbations.
However, such complication seems unavoidable since, on the one hand, non-separability is not unique to non-circular spacetimes, and, on the other hand, simple workarounds such as focusing on equatorial motion might be virtually uninformative in this respect.
Hence, although important work has been done on the subject already \cite{ChenTestingGravity2021,Chen:2023gwm,eichhorn_image_2021,eichhorn_locality-principle_2021,TakamoriTestingNoncircularity2023,GhoshParameterizedNoncircular2024,BabichevTestingDisformal2024a,GhoshProbingSpacetime2024}, the phenomenology of non-circular compact objects still remains largely to be explored.

\section*{Acknowledgement}

We wish to thank Aaron Held and Mokhtar Hassaine for the fruitful discussions.
The authors acknowledge support of ANR grant StronG (ANR-22-CE31-0015-01). 


\bibliographystyle{JHEP}
\bibliography{biblio.bib}

\end{document}